\RequirePackage{rotating}
\documentclass[a4paper,fleqn,usenatbib,useAMS]{mn2e}
\usepackage{rotating}
\usepackage{subfigure}
\RequirePackage{setspace}
\usepackage{graphicx}
\usepackage{fancyhdr}
\usepackage{natbib}
\usepackage{amsmath}
\usepackage{amssymb}
\usepackage{aas_macros}
\usepackage{epstopdf}
\usepackage{lscape}
\usepackage{lineno} 
\usepackage{color}

\title[GAMA/H-ATLAS: SFR indicators]{GAMA/H-ATLAS: A meta-analysis of SFR indicators - comprehensive measures of the SFR-M$_{*}$ relation and Cosmic Star Formation History at $z<0.4$}
\vspace{-5mm}
\author[L. J. M. Davies et. al.]{L. J. M. Davies$^{1}$\thanks{E-mail:
 luke.j.davies@uwa.edu.au}, S. P. Driver$^{1,2}$, A. S. G. Robotham$^{1}$, M. W. Grootes$^{3}$, \newauthor C. C. Popescu$^{5,6}$ R. J. Tuffs$^{4}$,  A. Hopkins$^{7}$, M. Alpaslan$^{8}$, S. K. Andrews$^{1}$, \newauthor J. Bland-Hawthorn$^{9}$, M. N. Bremer$^{10}$,    S. Brough$^{7}$,  M. J. I. Brown$^{11}$, M. E. Cluver$^{12}$,   \newauthor  S. Croom$^{9}$, E. da Cunha$^{13}$, L. Dunne$^{14,15}$, M. A. Lara-L\'opez$^{16}$, J. Liske$^{17}$, \newauthor J. Loveday$^{18}$,  A. J. Moffett$^{1}$,  M. Owers$^{7,19}$, S. Phillipps$^{10}$, A. E. Sansom$^{5}$, \newauthor E. N. Taylor$^{13}$,   M. J. Michalowski$^{14}$, E. Ibar$^{20}$,  M. Smith$^{15}$, N. Bourne$^{14}$ \\
 \\
$^{1}$ ICRAR, The University of Western Australia, 35 Stirling Highway, Crawley, WA 6009, Australia \\
$^{2}$ SUPA, School of Physics and Astronomy, University of St Andrews, North Haugh, St Andrews, Fife, KY16 9SS, UK\\
$^{3}$ ESA/ESTEC SCI-S, Keplerlaan 1, 2201 AZ Noordwijk, The Netherlands \\
$^{4}$ Max Planck Institut fuer Kernphysik, Saupfercheckweg 1, 69117 Heidelberg, Germany \\
$^{5}$ Jeremiah Horrocks Institute, University of Central Lancashire, Preston, UK\\
$^{6}$ The Astronomical Institute of the Romanian Academy, Str. Cutitul de Argint 5, Bucharest, Romania\\
$^{7}$ Australian Astronomical Observatory, North Ryde, P.O. Box 915, NSW 1670, Australia\\
$^{8}$ NASA Ames Research Centre, N232, Moffett Field, Mountain View, CA 94034, USA\\
$^{9}$ Sydney Institute for Astronomy, School of Physics A28, University of Sydney, NSW 2006, Australia.\\
$^{10}$ Astrophysics Group, School of Physics, University of Bristol, Tyndall Avenue, Bristol BS8 1TL\\
$^{11}$ School of Physics and Astronomy, Monash University, Clayton, Victoria 3800, Australia \\
$^{12}$ Department of Physics and Astronomy, university of the Western Cape, Robert Sobukwe Road, Bellville, 7535, South Africa\\
$^{13}$ Centre for Astrophysics \& Supercomputing, Swinburne University of Technology, PO Box 218, Hawthorn, VIC 3122, Australia\\
$^{14}$ Institute for Astronomy, University of Edinburgh, Royal Observatory, Edinburgh EH9 3HJ, UK\\
$^{15}$ School of Physics \& Astronomy, Cardiff University, The Parade, Cardiff CF24 3AA, UK\\
$^{16}$ Instituto de Astronom\'ia, Universidad Nacional Aut\'onoma de M\'exico, A.P. 70-264, 04510 M\'exico, D.F., M\'exico\\
$^{17}$ Hamburger Sternwarte, Universit{\'at} Hamburg, Gojenbergsweg 112, 21029 Hamburg, Germany\\
$^{18}$ Astronomy Centre, University of Sussex, Falmer, Brighton, BN1 9QH, UK \\
$^{19}$ Department of Physics and Astronomy, Macquarie University, NSW 2109, Australia \\
$^{20}$ Instituto de F\'isica y Astronom\'ia, Universidad de Valpara\'iso, Avda. Gran Breta\~na 1111, Valpara\'iso, Chile.}

\begin{document}

\date{Accepted: May 2016}

\pagerange{\pageref{firstpage}--\pageref{lastpage}} \pubyear{2016}

\maketitle

\vspace{-3mm}

\begin{abstract}

\vspace{-3mm}

We present a meta-analysis of star-formation rate (SFR) indicators in the GAMA survey, producing 12 different SFR metrics and determining the SFR-M$_{*}$ relation for each. We compare and contrast published methods to extract the SFR from each indicator, using a well-defined local sample of morphologically-selected spiral galaxies, which
excludes sources which potentially have large recent changes to their SFR. The different methods are found to yield
SFR-M$_{*}$ relations with inconsistent slopes and normalisations, suggesting differences between calibration methods. The recovered SFR-M$_{*}$ relations also have a large range in scatter which, as SFRs of the targets may be considered constant over the different timescales, suggests differences in the accuracy by which methods correct for attenuation in individual targets. We then recalibrate all SFR indicators to provide new, robust and consistent luminosity-to-SFR calibrations, finding that the most consistent slopes and normalisations of the SFR-M$_{*}$ relations are obtained when recalibrated using the radiation transfer method of Popescu et al. These new calibrations can be used to directly compare SFRs across different observations, epochs and galaxy populations. We then apply our calibrations to the GAMA II equatorial dataset and explore the evolution of star-formation in the local Universe. We determine the evolution of the normalisation to the SFR-M$_{*}$ relation from $0<z<0.35$ - finding consistent trends with previous estimates at $0.3<z<1.2$. We then provide the definitive $z<0.35$ Cosmic Star Formation History, SFR-M$_{*}$ relation and its evolution over the last 3\,billion years.

\end{abstract}


\section{Introduction}
\label{sec:Intro}

Measuring the rate at which new stars are forming in galaxies both in the local Universe and as a function of time is crucial to our understanding of the initial formation and subsequent evolution of galaxies. Galaxies form from over-dense regions in clouds of molecular and atomic gas \citep[$e.g.$ see][]{White91, Kauffmann93}, and then grow and evolve via both the formation of new stars \citep[star formation, $e.g.$ see summary in][]{Kennicutt98b} and hierarchical merging over cosmic timescales \citep[mergers, $e.g.$ see summary in][]{Baugh06}. The former is intimately linked to internal physical characteristics of a specific galaxy: atomic and molecular gas mass and distribution \citep{Kennicutt98b, Keres05}, stellar mass \citep[$e.g.$][]{Brinchmann04, Noeske07, Daddi07,Elbaz07}, dust mass \citep{DaCunha10}, morphology \citep[$e.g.$][]{Kauffmann03, Guglielmo15}, active galactic nuclei (AGN) activity \citep{Netzer09, Thacker14} and metallicity \citep{Ellison08, Mannucci10,Lara-Lopez13}. In contrast the latter (galaxy mergers) is fundamentally linked to external characteristics of the local environment, such as a galaxy's local mass density - $i.e.$ the mass and distribution of matter in the galaxy's local environment: pairs, groups, clusters, etc \citep[$e.g.$][]{McIntosh08, Ellison10, DeRavel}.       

However, these processes are not mutually exclusive as local galaxy interactions \citep[$e.g.$][]{Scudder12,Robotham14,Davies15b,Davies16} and large scale environment \citep{Peng10} can have a strong effect on a galaxy's growth via star-formation. In order to gain a clear picture of both the fundamental process of how galaxies grow in mass via forming new stars and the effects of local environment on these processes, we must accurately measure the current star formation rate (SFR) over a seemingly disparate range of galaxy types, environments and cosmic timescales. 

A key diagnostic of both the distribution and evolution of star formation in the Universe is the so-called star-forming galaxy main sequence \citep[the SFR-M$_{*}$ relation, $e.g.$][]{Brinchmann04, Salim07, Gilbank11, Whitaker12, Bauer13, Grootes14, Lee15,Tomczak16}. This relation displays the tight correlation between stellar mass and star formation in actively star-forming galaxies and appears to evolve in normalisation out to high redshift \citep[$e.g.$][]{Noeske07, Daddi07,Elbaz07,Lee15}. The ubiquity of this relation over an extensive redshift range \citep[$e.g.$][]{Tomczak16} and its relatively small scatter \citep[$e.g.$][]{Salim07,Guo13}, make it a powerful tool in determining the changing rate at which galaxies of a fixed mass form stars as the Universe evolves \citep[$e.g.$][]{Lee15}. The consensus interpretation for the physical processes underlying this relation, is that star-formation in the bulk of galaxies is produced by relatively stable processes, such as gas infall and accretion. 

This evolution in star formation is also traced by the global distribution of star formation in the Universe per unit volume, as a function of redshift \citep[the Cosmic Star-Formation History, CSFH, $e.g.$][ - and see review in Madau \& Dickinson,  2014]{Lily96, Madau96, Baldry02, Hopkins06,Behroozi13}. This represents a fundamental measurement of the process of galaxy evolution, probing the changing density of star-formation as the Universe evolves \citep[$e.g.$][]{Baldry02}. As such, this distribution contains invaluable information about the underlying processes which shape the evolution of galaxies. The CSFH has been well established over the past two decades from numerous studies, albeit with large scatter from different data sets \citep[see summaries in][]{Hopkins06,Gunawardhana13} - showing an increase in SFR density to $z\sim2.5$ and a slow decrease at later times. This distribution traces the initial formation of galaxies, their rapid growth via the conversion of neutral gas into stars and the slow decline of global star-formation.

In order to robustly probe the evolution of star-formation via both the SFR-M$_{*}$ relation and the CSFH we need consistent measurements of star-formation across an extensive ($\sim13$\,Gyr) evolutionary baseline. However, this is problematic as the key observables required to determine the rate at which a galaxy is forming stars are specific to galaxy populations, the underlying physical processes occurring within a galaxy (most notably the absorption and scattering of starlight by interstellar dust grains) and the epoch at which the galaxies reside. For example, H-$\alpha$ emission is redshifted out of the optical bands at $z>0.5$ and FIR observations become highly spatially confused at depths probing relatively modest SFRs outside of the local Universe. This variation in SFR measurements makes comparisons across multiple surveys, observations and epochs problematic. 

However, it is possible to measure SFRs using multiple methods over a consistent and statistically robust sample of galaxies, which are selected to have relatively steady-state SFRs ($i.e.$ not subject to recent changes to their star-formation history) and calibrate all methods such that they produce the same measured SFR for the general population. These calibrations can then be applied to give consistent results across observations which probe fundamentally different SFR tracers. Once robustly calibrated, we can explore differences in SFRs derived using different methods as being representative of systems with non-steady-state SFRs. For example, once calibrated to the same baseline for the general population, we can explore variations between long-duration and short-duration star-formation methods ($i.e.$ FIR to H-$\alpha$) as being representative of recent changes to a galaxy's star-formation history \citep[$e.g.$ see][]{Davies15b}.

An ideal sample for this analysis is the Galaxy And Mass Assembly (GAMA) survey. GAMA is a highly complete multi-wavelength database \citep{Driver11, Driver15} and galaxy redshift ($z$) survey \citep{Baldry10,Hopkins13, Liske15} covering $\sim$280\,deg$^{2}$ and containing $\sim250,000$ spectroscopically confirmed sources. The GAMA data consist of 21-band photometric data (for UV, optical, MIR and FIR SF indicators), spectroscopic line emission measurements (for [OII] and H$\alpha$ SF indicators) and multiple studies which can be used to derive SFR measurements (such as radiative transfer dust modelling to correct UV fluxes and full SED fits), for all $\sim250,000$ galaxies.

In this work we use the extensive GAMA database to derive SFRs across all possible available methods. We use the SFR-M$_{*}$ relation as a fundamental diagnostic tool, and compare and contrast the current literature SFR calibrations for all methods. We then use SFRs derived from all methods to produce new luminosity to SFR calibrations. We propose new robust and consistent calibrations which can be used to derive consistent SFRs across all methods and as such, can be used to compare SFRs across diverse galaxy populations, environments, epochs and surveys. We then apply our newly calibrated SFRs to the full GAMA sample and investigate the evolution of the SFR-M$_{*}$ relation and CSFH. 

We refer the casual reader to our final Luminosity-SFR calibrations for the GAMA sample, which can be used to consistently measure SFR across multiple observables - outlined in Table \ref{tab:pars}, and our discussion of the evolution of the SFR-M$_{*}$ relation and CSFH - in Section \ref{sec:dis}.

Throughout this paper we use a standard $\Lambda$CDM cosmology with {H}$_{0}$\,=\,70\,kms$^{-1}$\,Mpc$^{-1}$, $\Omega_{\Lambda}$\,=\,0.7 and $\Omega_{M}$\,=\,0.3.

\begin{figure}
\includegraphics[scale=0.6]{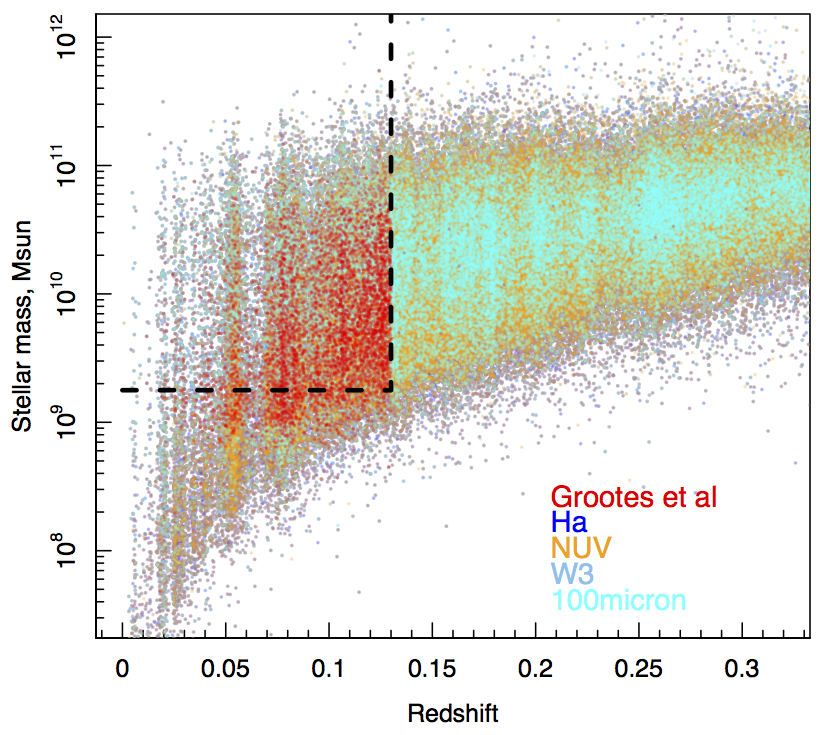}
\caption{The stellar mass distribution of GAMA II galaxies as a function of redshift. Points are coloured by observation band/emission line feature in which they are detected at $>2\sigma$. The sample used in this work is bounded by the dashed box. Points are overlaid in colour, such that galaxies which are detected at multiple wavelengths only display one colours. This figure is intended to show that the Grootes et al sample is volume limited and covers detections over a range of our SFR methods. Note that $>2\sigma$ cuts are not applied to our sample.}
\label{fig:mass_v_z}
\end{figure}
    
\section{Data}
\label{sec:data}

The GAMA survey  second data release (GAMA II) covers 286\,deg$^{2}$ to a main survey limit of $r_{\mathrm{AB}}<19.8$\,mag in three equatorial (G09, G12 and G15) and two southern (G02 and G23 \- survey limit of $i_{\mathrm{AB}}<19.2$\,mag in G23) regions. The limiting magnitude of GAMA was initially designed to probe all aspects of cosmic structures on 1\,kpc to 1\,Mpc scales spanning all environments and out to a redshift limit of $z\sim$0.4. The spectroscopic survey was undertaken using the AAOmega fibre-fed spectrograph \citep[][]{Sharp06,Saunders04} in conjunction with the Two-degree Field \citep[2dF,][]{Lewis02} positioner on the Anglo-Australian Telescope and obtained redshifts for $\sim$240,000 targets covering $0<z\lesssim0.5$ with a median redshift of $z\sim0.2$, and highly uniform spatial completeness \citep[see][for summary of GAMA observations]{Baldry10,Robotham10,Driver11}. 

Full details of the GAMA survey can be found in \citet{Driver11, Driver16} and \citet{Liske15}. In this work we utilise the data obtained in the 3 equatorial regions, which we refer to here as GAMA II$_{Eq}$. Stellar masses for the GAMA II$_{Eq}$ sample are derived from the $ugriZYJHK$ photometry using a method similar to that outlined in \cite{Taylor11} - assuming a Chabrier IMF \citep{Chabrier03}.

As we aim to derive new SFR calibrations across multiple SFR methods, we require a well defined and constrained sample of star-forming galaxies.

Therefore, we use the low contamination and high completeness, volume limited sample of spiral galaxies outlined in Grootes et al (submitted) selected following the method presented in \cite{Grootes14}. This sample and the method are defined in detail in these works. Briefly, the sample uses a non-parametric, cell-based, morphological classification algorithm to identify spiral galaxies at $0<z<0.13$. Through subsequent visual inspection, the algorithm has been found to successfully identify such systems over this epoch and mass range. The morphological proxy parameters used in Grootes et al. are the $r$-band effective radius, $i$-band luminosity and single-sersic index, importantly avoiding observables which are themselves SFR indicators. We refer the reader to \cite{Grootes14} and Grootes et al. (submitted) for further details. We also apply an initial lower stellar mass limit selection of $log_{10}[M_{*}/M_{\odot}]>9.25$ to avoid stellar masses where the Grootes et al. sample is likely to become incomplete.   

Our sample selection in comparison to a number of our SFR tracers (outlined in the following section) is shown in Figure \ref{fig:mass_v_z}. We display all sources which are detected at $>2\sigma$ in H$\alpha$, NUV, W3 and 100$\mu$m. Note that we do not apply these $>2\sigma$ cuts to our sample, they are purely used to display the distribution of sources over a sample of SFR methods. Red points display the Grootes et al sample, and the black dashed region displays our selection window. The \cite{Grootes14} sample, with the above stellar mass cut, gives 5,311 sources. We then also remove sources which potentially contain an AGN, identified using their emission line diagnostics as in \cite{Davies15b}, and exclude sources identified to be in a pair/group using the GAMA group catalog \citep[for further details see][]{Robotham11}, leaving 3,749 galaxies. \cite{Davies15b,Davies16} and Grootes et al (submitted) find that the pair and group environment can strongly affect star formation in galaxies, even in non-interacting galaxies, and that these changes can occur on short timescales which induce scatter in the SFR-M$_{*}$ relation ($e.g.$ Grootes et al., submitted). Given that the SFR methods used in this work probe fundamentally different timescales, galaxies which have undergone recent changes to their star-formation, will have intrinsically different SFR when measured using different methods \citep[$e.g.$ see][]{Davies15b}. By excluding pair/group galaxies, we aim to remove sources which have had recent changes to their star formation, and thus aim to restrict our sample to galaxies which have intrinsically the same SFR, irrespective of the timescale a particular SFR method probes. While there will still be sources within our sample that have undergone recent changes to their star formation (in particular remaining members of pairs where the other pair galaxy falls below our survey limit and thus the system is not identified as such), the general trends of the population should display consistent intrinsic SFRs. In fact, once calibrated to each other to remove systematic offsets (as we do here), scatter between SFRs from different indicators may potentially be a sign of short timescale variations in star formation history. To reveal such temporal variations would also require a high fidelity for correction of dust attenuation on an object-by-object basis, due to the very small intrinsic scatter of individual galaxies in the SFR-M$_{*}$ relation for the spiral galaxies used here.

Note that we do not exclude sources which are undetected in a particular indicator, but use the full sample defined above for our re-calibrations. The GAMA analysis provides a measurement for all indicators/methods for every source in the Grootes et al sample, albeit some which are consistent with zero, but with realistic errors. As such, some indicators ($i.e.$ those using MIR-FIR data) will have derived SFRs with large errors (and consistent with zero SFR) at the low luminosity end. As detailed in the description of the fitting process in
Section \ref{sec:com_ind}, we address this issue by raising the lower limit in stellar mass in the shallowest indicators (W4 and 100$\mu$m). This ensures that noise-dominated data never dominates the statistical analysis in any part of the stellar mass range. 

Finally, it is important to note that the final sample encompasses a full range of objects with differing dust content and dust opacities. Thus it avoids bias in inter-comparing methods for determining SFR from dust-reradiated starlight with methods from direct measurements of starlight. Such bias would arise if one were to select the sample for statistical comparison according to detectability in the SFR indicators themselves, especially where disk opacity is a strong function of stellar mass and luminosity.

\section{Methods for deriving SFRs}
\label{sec:indicators}

In this work we use multiple methods for determining SFRs and derive new luminosity-SFR calibrations. Initially we explore all SFR methods available in the GAMA II$_{Eq}$ sample, using previously published (or in preparation) GAMA SFR calibrations. While there is a wide breadth of varying SFR calibrations for each observable in the literature \citep[$e.g.$][]{Kennicutt12}, we limit ourselves to those used previously in GAMA, for the sake of clarity. Below we detail each method for determining SFRs, the observable emission from which it is derived, and the physical process from which the emission is produced. Ultimately, all the methods - irrespective of whether the observable is directly starlight, dust-reradiated starlight, or nebular line emission - quantitatively link the observable to the intrinsic emission from hot stars ($i.e.$ as would be observed in the absence of dust). A major part of each method is therefore concerned with how the corrections for dust attenuation are made.

In this section, unless otherwise stated, we use photometry measured by the Lambda Adaptive Multi-Band Deblending Algorithm for R  \citep[\textsc{lambdar}][]{Wright16}, with rest-frame, k-corrected measurements derived from a refactored implementation of the \textsc{interrest} algorithm \citep{Taylor09}, coupled with the empirical set of galaxy template spectra of \cite{Brown14}. We also apply simple linear scalings, which are appropriate for young ($<$1\,Gyr), high mass stars to ensure all SFRs are consistent with an assumed Chabrier IMF \citep[$e.g.$ see Table 1 of ][]{Driver13}, and see Figure 3 of \cite{Courteau14} for a comparison of IMFs. Potentially there are also differences in methods due to the assumed stellar population synthesis (SPS) models used to derive SFRs using a particular method \cite[$e.g.$ see][for a comparison of H$\alpha$ and NUV SFRs, where differences between SFRs are attributed to the SPS models used]{Salim07}. However, this variation is small and will be removed in our re-calibration process.

We define all of the SFR methods described below as `base' calibrations, and refer to them as such in the text and figures.

\subsection{H$\alpha$}
\label{sec:Ha}

H$\alpha$ photons arise from gas ionised by the stellar radiation field, and only stars with ages $<$20\,Myr can contribute significantly to this ionizing flux. Thus, H$\alpha$ provides a direct measure of the current SFR in galaxies ($<$10-20\,Myr) which is largely independent of SF history \citep[$e.g.$ see][]{Kennicutt98a}. 

For SFR$_{\mathrm{H\alpha}}$ we use emission line data from the GAMA II$_{Eq}$ spectroscopic campaign, where aperture, obscuration and stellar absorption corrected H$\alpha$ luminosities following \cite{Hopkins03}:                 

\begin{equation}
\label{eq:Ha_lum}
\begin{split}
\mathrm{L_{H\alpha}} &= \mathrm{(EW_{H\alpha} + EW_{c})} \times 10^{-0.4(M_{r}-34.1)} \\
                       & \ \ \ \  \mathrm{\times \frac{3\times10^{18}}{(6564.1(1+z)^2)} \left(\frac{F_{H\alpha}/F_{H\beta}}{2.86} \right) ^{2.36}},
\end{split}
\end{equation}

\noindent and $\mathrm{EW_{H\alpha}} $ denotes the H$\alpha$ equivalent width, $\mathrm{EW_c}$ is the equivalent width correction for stellar absorption \citep[2.5\AA\ for GAMA,][]{Hopkins13}, $M_{r}$ is the galaxy $r$-band rest-frame absolute magnitude and $\mathrm{F_{H\alpha}/F_{H\beta}}$ is the Balmer decrement \citep[see][for further details]{Gunawardhana11}. While the individual stellar absorption corrections will depend on star formation history, we opt to use a single value as in \cite{Gunawardhana11,Gunawardhana15} and \cite{Hopkins13} - please refer to these papers for further details.
One caveat to using H$\alpha$ as a star-formation rate indicator is that the aperture based spectroscopy only probes the central regions of nearby galaxies. However, the equation above aims to correct for this effect by applying a scaling based on the absolute $r$-band magnitude. Using this, SFR$_{\mathrm{H\alpha}}$ can be determined from \cite{Kennicutt98a}, which assuming a Salpeter IMF, with a conversion to Chabrier IMF using Table 1 of \cite{Driver13}:

\begin{equation}
\mathrm{SFR_{\mathrm{H\alpha}} (M_{\odot}/yr)=\frac{L_{H\alpha} (W\,Hz^{-1})}{1.27 \times 10^{34} } \times 1.53 }.
\end{equation}

Errors for SFR$_{\mathrm{H\alpha}}$ are estimated by propagating the measurement errors associated with the EW, flux and $r$-band magnitude in Equation \ref{eq:Ha_lum}.

\subsection{[OII]}

[OII] emission arises from similar nebular regions as H$\alpha$, and as such, can also be used to trace short timescale ($<$20\,Myr) SF in galaxies \citep[e.g.][]{Gallagher89, Kennicutt98a, Kewley04}. Using the [OII] emission lines to derive SFRs is especially important at intermediate-high redshift ($z>0.5$), as the H$\alpha$ emission line feature moves from optical to NIR wavelengths. However, accurately measuring [OII] luminosities, with which to derive SFRs is fraught with difficulty due to strong dependance on dust reddening and oxygen abundance \citep[see][and references therein]{Kewley04}.           

Here, we derive [OII] based SFRs in a similar manner to H$\alpha$ and following the prescription for GAMA, originally outlined in \cite{Hopkins03} and \cite{Wijesinghe11}. We once again use GAMA emission line data, where aperture and obscuration corrected [OII] luminosities are given by:   

\begin{equation}
\label{eq:OII_lum_a}
\begin{split}
\mathrm{L_{[OII]Obs}} &= \mathrm{(EW_{[OII]})} \times 10^{-0.4(M_{u}-34.1)} \\
                       & \ \ \ \  \mathrm{\times \frac{3\times10^{18}}{(3728.3(1+z)^2)} \left(\frac{F_{H\alpha}/F_{H\beta}}{2.86} \right)^{2.36}}.
\end{split}
\end{equation}

Note that we do not apply a stellar absorption correction (as it is not required) to the [OII] emission as we do with H$\alpha$. Unlike H$\alpha$, [OII] luminosity to SFRs calibrations can be strongly affected by metallicity \cite[$e.g.$][]{Kewley04}. As such, we apply a metallicity-dependent correction to the [OII] luminosity following \cite{Kewley04}, where: 

\begin{equation}
\label{eq:OII_lum}
\mathrm{L_{[OII]}} = \mathrm{\frac{L_{[OII]Obs}}{-1.75[log(O/H)+12]+16.73}}
\end{equation}

\noindent We do not measure log(O/H)+12 for individual systems, as these metallicity measurements may induce large correction errors, but instead use the  log(O/H)+12 to stellar mass relation from Eq. 7 of \cite{Lara-Lopez13}:

\begin{equation}
\label{eq:metal}
\mathrm{log(O/H)+12} = \mathrm{-10.83+3.65log[M]-0.17log[M]^2}
\end{equation}

\noindent In this manner we essentially apply a stellar mass-dependant weighting to our [OII] luminosities. We then use the [OII] luminosity to SFR calibration of \cite{Wijesinghe11}, with a conversion from \cite{Baldry03} IMF to Chabrier IMF: 

\begin{equation}
\mathrm{SFR_{\mathrm{[OII]}} (M_{\odot}/yr)=\frac{L_{[OII]} (W\,Hz^{-1})}{7.97 \times 10^{33} }\times 0.824}.
\end{equation}

Errors for SFR$_{\mathrm{[OII]}}$ are also estimated by propagating the measurement errors associated with the EW, flux and $u$-band magnitude in Equation \ref{eq:OII_lum_a}.

\subsection{$\beta-$corrected GALEX-NUV and -FUV}
\label{sec:UV}

UV continuum emission arises from hot, massive (M$_{*} >$ 3\,M$_{\odot}$) O and B stars, and as such is a good tracer of more recent SF in galaxies \citep[$e.g.$][]{Kennicutt12}, with luminosity-weighted mean ages for a constant SFR predicted to be $\sim$28\,Myr for the GALEX FUV band, and $\sim$80\,Myr for GALEX NUV (Grootes et al. submitted). However, this emission is easily obscured by dust, which we must correct for to varying degrees depending on source type and inclination. In the most simplistic treatment of UV measurements, SFRs can be derived by correcting the observed GALEX-NUV and -FUV luminosity using a obscuration correction derived from the UV spectral slope ($\beta$).  This method assumes an intrinsic UV spectral slope, and attributes any deviation from this slope to be caused by dust obscuration \citep[see][]{Meurer99}. We use \textsc{lambdar}-derived photometry, and calculate the UV spectral slope as:

\begin{equation}
\mathrm{\beta_{UV}=\dfrac{log_{10}[L_{1528Obs}]-log_{10}[L_{2271Obs}]}{log_{10}[1528]-log_{10}[2271]}   },
\end{equation}

we then obscuration-correct the observed both UV luminosities as:

\begin{equation}
\label{eq:UV_lum}
\mathrm{L_{UV} (W\,Hz^{-1}) = L_{UVObs} (W\,Hz^{-1})10^{0.4(4.43+1.99\beta_{UV})}}
\end{equation}

and derive, UV SFRs using the \cite{Salim07} relation for a Chabrier IMF:

\begin{equation}
\mathrm{SFR_{\mathrm{NUV}} (M_{\odot}/yr)=\dfrac{L_{2271} (W\,Hz^{-1})}{1.38\times10^{21}} }
\end{equation}

\begin{equation}
\mathrm{SFR_{\mathrm{FUV}} (M_{\odot}/yr)=\dfrac{L_{1528} (W\,Hz^{-1})}{1.46\times10^{21}} }
\end{equation}

Errors for UV-based SFR are estimated by propagating the measurement errors associated with the NUV and FUV \textsc{lambdar}-derived photometry \cite[see][for details of \textsc{lambdar} errors]{Wright16}.

\subsection{$u$-band}

Following \cite{Hopkins03} we also consider SFRs derived using the rest-frame $u$-band ($\sim3500$\AA) emission. Flux observed at these wavelengths is thought to arise from the photospheres of young, massive stars and as such, traces star-formation on $\lesssim100$\,Myr timescales. This emission is less strongly affected by dust obscuration than the UV, but has a more tenuous link to the emission arising purely from hot, young stars as it has a large contribution from older stellar populations, see below. Here we use k-corrected and obscuration-corrected rest-frame $u$-band photometry from \textsc{lambdar}/\textsc{interrest}. 

Briefly, \textsc{interrest} is used to fit the \textsc{lambdar} photometry using the empirical stellar population template set from \cite{Brown14}, with a single-screen \cite{Calzetti00} dust attenuation law applied, where the degree of attenuation is characterised by the selective extinction between the B- and V-bands, E(B-V). The best fitting template is then used to derive the un-attenuated, rest-frame $u$-band photometry \citep[see][for further details of a similar process]{Taylor11}. As such, our $u$-band SFRs contain the implicit assumption that a \cite{Calzetti00} extinction law is appropriate for all sources and that the \cite{Brown14} templates cover the intrinsic properties of the full range of galaxies in our sample. We then apply the $u$-band luminosity to SFR calibration derived in \cite{Hopkins03}, with a conversion from Salpeter IMF to Chabrier IMF:

\begin{equation}
\label{eq:u}
\mathrm{SFR_{u} (M_{\odot}/yr)=\left( \frac{L_{uObs} (W\,Hz^{-1})}{1.81 \times 10^{21}} \right)^{1.186} \times 1.53}.
\end{equation}
  
We note that the \cite{Hopkins03} relation is calibrated to the H$\alpha$ measurements of SDSS galaxies, which in turn are based on the the same SFR method as used in Section \ref{sec:Ha}. This $u$-band SFR calibration contains a non-linear term which essentially accounts for the fact that, as noted above, not all $u$-band emission arises from young stars ($i.e.$ there is a fraction of $u$-band emission that arises from older stellar populations). This will be discussed further in subsequent sections. However, in the recalibrations we use in this paper, we only derive linear relations between luminosity and SFR. Thus, we initially correct the $u$-band luminosity given in Eq. \ref{eq:u} to account for contributions from old stars. To do this we apply a colour dependant (as a proxy for stellar mass) scaling to the k-corrected and obscuration-corrected $u$-band luminosity as:

\begin{equation}
\label{eq:u_lum}
\mathrm{L_{u} (W\,Hz^{-1}) = L_{uObs} (W\,Hz^{-1}) \times (-1.1(u-g)+1.6)}.
\end{equation}  

\noindent for sources with $u-g>$0.55. At $u-g\lesssim$0.55 (the very bluest galaxies) we assume all emission in the $u$-band arrises from star formation. This scaling is estimated from the fraction of $u$-band emission arising star formation given in \cite{Kennicutt98b} - see their Figure 2. Applying this scaling is found to remove non-linearity in L$_{u}$ to SFR relations. Errors for $u$-band SFR are estimated from the propagated \textsc{interrest}-derived rest-frame photometry errors. In \textsc{interrest}, these errors are calculated from the formal uncertainty of the stellar population fits at a specific waveband.

\subsection{100$\mu$m and WISE W3/W4}

Infrared (IR) emission at 1-1000$\mu$m arising from `normal' star-forming galaxies is produced from three sources: photospheres and circumstellar envelopes of old stars undergoing mass loss \citep[$e.g.$][]{Melbourne12}, interstellar gas and dust heated by either bright OB stars in star-forming regions (warm dust) or the general stellar radiation field throughout the interstellar medium (ISM, cool dust - `cirrus') - for example see the review by \cite{Sauvage05}, \cite{Popescu00} or more recently \cite{Xilouris12} and references therein. Stellar sources of IR emission dominate at short wavelengths $<3\mu$m and interstellar gas emission makes up just a few percent of the total IR output of galaxies. At 22-100$\mu$m the bulk of the emission arises from warm dust locally
heated by UV emission from young stars in star-forming regions, with cirrus
emission dominating at $>$100 $\mu$m and between 3-22$\mu$m \cite[see][]{Popescu11}. As such, probing IR flux from star-forming galaxies in the $3 - 100\mu$m range gives a reliable estimate for the ongoing SF \citep[$e.g.$][the amount of flux emitted in the IR is directly related to the UV emission from newly formed stars]{Calzetti07}. However, IR emission from dust requires significantly long timescales to become apparent and subsides slowly when SF is suppressed \citep[see][]{Kennicutt98a}. 

In this paper we consider three IR continuum measures of SFR. Firstly, we determine the SFRs measured from the \textsc{lambdar}-derived 100$\mu$m flux. The 100$\mu$m data was provided to GAMA as part of the Herschel Astrophysical Terahertz Large Area Survey \citep[H-ATLAS][]{Eales10}. The H-ATLAS 100$\mu$m data will be published in Valiante, in prep following updated data reduction from \cite{Ibar10}. We convert 100$\mu$m fluxes to SFRs using the tight correlation derived in \cite{Davies14} for Virgo cluster galaxies:

\begin{equation}
\mathrm{log_{10}\,SFR_{p100}(M_{\sun}yr^{-1}) = 0.73\,log_{10}L_{100}(W\,Hz^{-1})- 17.1 },    
\end{equation}

\noindent which compares 100$\mu$m emission to UV-derived SFRs using the method of \cite{Iglesias-Paramo06}.

Secondly, we estimate SFRs using \textsc{lambdar}/\textsc{interrest} photometry derived from the $Wide-field Infrared Survey Explorer$ ($WISE$) data outlined in \cite{Cluver14}. We use $WISE$ 12$\mu$m (W3) and $22\mu$m (W4) band fluxes, and the best-fit SFR correlations obtained in \cite{Cluver14}, with a conversion from Salpeter IMF to Chabrier IMF:    

\begin{equation}
\mathrm{SFR_{\mathrm{W3}} (M_{\odot}\,yr^{-1})=10^{1.13\,\mathrm{log}_{10} \nu_{12\mu m} L_{12\mu m}(L_{\odot})-10.24} \times 1.53}.
\end{equation}

\begin{equation}
\mathrm{SFR_{\mathrm{W4}} (M_{\odot}\,yr^{-1})=10.^{0.82\,\mathrm{log}_{10} \nu_{22\mu m} L_{22\mu m}(L_{\odot})-7.3} \times 1.53}.
\end{equation}

Note that these calibrations are also based on the H$\alpha$ derived SFRs for GAMA galaxies from \cite{Gunawardhana11}, and thus are subject to the same assumptions as in Section \ref{sec:Ha}. Errors in the MIR/FIR are estimated by propagating the measurement errors associated with the \textsc{lambdar}-derived photometry \cite[see][for details of \textsc{lambdar} errors]{Wright16}.

\begin{figure*}
\begin{center}
\includegraphics[scale=0.75]{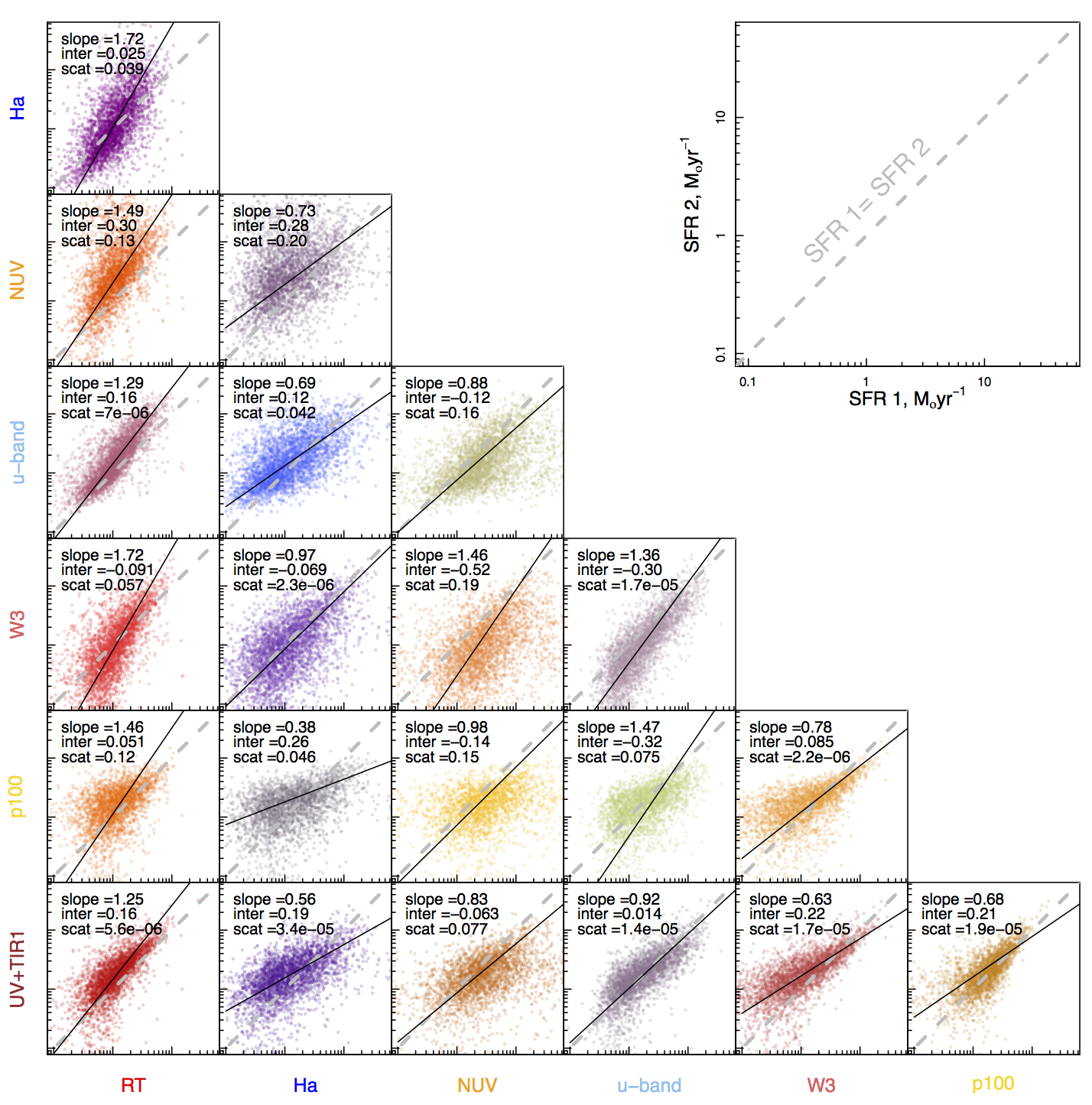}
\caption{A comparison of selected SFR indicators for the literature SFR calibrations described in Section \ref{sec:indicators}. Both axes display SFR from 0.1 to 50\,M$_{\odot}$\,yr$^{-1}$ in log-log space and the grey central dashed line displays a 1:1 relation - shown in the offset panel. In each panel we display the slope, intercept and orthogonal intrinsic scatter (note this is the scatter not accounted for by error in the measurement - $i.e.$ a very small scatter implies that the intrinsic scatter in the distribution is much smaller then the measurement error) of the best linear fit. Throughout this paper we assign a specific colour to each SFR indicator (shown in the axes labels). The colours in this figure represent a combination of the colours assigned to the indicators compared in each panel. Note that the statistical comparison was done on the full optically-defined sample selected as defined in Section \ref{sec:data}. However, not all points can be displayed due to negative flux measurement in the plotted SFR indicator (see text for details). }

\label{fig:SF_vs_SF_old}
\end{center}
\end{figure*}

\subsection{UV+TIR1}

We also use the combination of UV and total IR (TIR) luminosities as a star formation rate proxy broadly probing the last $\sim300$\,Myr of the galaxies star-formation history. As discussed above, UV emission arises directly from star-forming regions and probes star-formation on short timescales, while some fraction of this emission is absorbed and reprocessed by dust, being re-emitted in the FIR on longer timescales. As such, using a star-formation rate determination method which probes both the UV and FIR emission will give a relatively stable, but broad timescale, measure of star-formation in our sample galaxies.  We therefore sum both UV and TIR luminosities to obtain a total star formation rate estimate, based on the bolometric luminosity of OB stars. This is analogous to correcting the UV emission for the fraction of light absorbed and re-emitted in the FIR. 

We use the method outlined in many high redshift studies \citep[e.g.][]{Bell05,Papovich07, Barro11} of:

\begin{equation}
\label{eq:UVTIR1_lum}
\mathrm{SFR_{UV+TIR1}(M_{\odot}\,yr^{-1}) = 1.09 \times 10^{-10}[L_{IR}+ 2.2 L_{UV}](L_{\odot}) }.
\end{equation}

This prescription is the \cite{Bell05} recalibration of the relation from \cite{Kennicutt98a}, scaled for a \cite{Chabrier03} stellar IMF. 

Here we use the \citet{Brown14} spectrophotometrically calibrated library of
galaxy spectra to derive UV and TIR luminosities. We follow a Bayesian process,
with uniform/uninformative priors on the templates ($i.e.$ each template is
assumed to be equally likely). For a particular template, the best fit/maximum
likelihood value and the formal uncertainty are analytic (through the
usual propagation of uncertainties). The posterior for the best-fit value
template is given by marginalising over the full set of templates. By effectively
marginalising over template number as a nuisance parameter, we fully
propagated the errors, including uncertainties due to template ambiguities. 

Note that the photometry used in this case is from the GAMA II panchromatic data release \citep{Driver15} and not that derived from \textsc{lambdar}.

\subsection{UV+TIR2}

In a similar approach to UV+TIR1, Wang et al (in prep) also calculate UV + total IR (TIR) SFRs for the GAMA II$_{Eq}$ sample. They use the GALEX-NUV measurements and combine with independent fits to the FIR data using the \cite{Elbaz01} templates. Final SFRs are calculated using: 

\begin{equation}
\mathrm{SFR_{UV+TIR2}(M_{\odot}\,yr^{-1}) = 1.72 \times 10^{-10}[L_{IR}+ 0.47L_{UV}](L_{\odot}) }.
\end{equation}

Note that once again the photometry used in this case is from the GAMA II panchromatic data release \citep{Driver15} and not that derived from \textsc{lambdar}. For further details of this process, including how the UV and IR measurements were obtained, please refer to Wang et al. (in prep).

For our luminosity to SFR calibrations we opt to use the same luminosity form as UV+TIR1. Hence, in the rest of this paper we define the following luminosity, using the L$_{IR}$ and L$_{UV}$ measurements outlined in Wang et al. (in prep):

\begin{equation}
\label{eq:UVTIR2_lum}
\mathrm{L_{UV+TIR2} = L_{IR}+ 2.2L_{UV}}.
\end{equation}

The errors on SFR$_{UV+TIR2}$ are also calculated by a propagation of the combination of the UV and TIR errors. UV errors are once again derived from photometric errors from the GAMA II panchromatic data release and TIR errors are derived by marginalising over the effective dust temperature in the \cite{Elbaz01} template fits (see Wang et al. in prep for further details).

\subsection{MAGPHYS}

Spectral Energy Distribution (SED) fitting codes model the full UV-to-IR spectrum of a galaxy in a physical way and estimate recent star-formation. Here we use the full SED \textsc{magphys} \citep{daCunha08} fits to the GAMA II galaxies (Driver et al in prep). In this upcoming work, Driver et al obtain the best SED fits to the full 21-band photometric data available for all GAMA II sources, simultaneously fitting the UV through FIR flux. The \textsc{magphys} code provides an estimate of the galaxy SFR averaged over the last 100\,Myr using a best fit energy balance model - where the obscuration corrected SED is determined by balancing energy absorbed in the UV/optical with that emitted in the IR for a set of physically plausible priors for SFH. As such, the \textsc{magphys} method applies a physically meaningful, SED derived, obscuration correction to determine SFRs (hereafter SFR$_{\textsc{magphys}}$). 

Briefly, \textsc{magphys} uses the \cite{Bruzual03} stellar populations with a \cite{Chabrier03} IMF and assumes an angle-averaged attenuation model of \cite{Charlot00}. This is combined with an empirical NIR-FIR model accounting for PAH features and near-IR continuum emission, emission from hot dust and emission from thermal dust in equilibrium. The code defines a model library over a wide range of star formation histories, metallicities, and dust masses and temperatures, and fits the \textsc{lambdar} photometry - forcing energy balance between the observed TIR emission and the obscured flux in the UV-optical. Physical properties (SFR, SFH, metallicity, dust mass, dust temperature) for the galaxy are then estimated from the model fits, giving various percentile ranges for each parameter. Here we use the median SFR$_{0.1Gyr}$ parameter, which provides an estimate for the SFR averaged over the last 0.1\,Gyrs. Errors on SFR$_{\textsc{magphys}}$ are estimated from the 16th-84th percentile range of the SFR$_{0.1Gyr}$ parameter, which encompasses both measurement and fitting errors.

\subsection{Radiative Transfer dust-corrected NUV}

A more complex method which provides an attenuation corrected SFR measurements derived
self-consistently using information from the whole range of the electromagnetic
spectrum, from the UV to the FIR/submm, is
the radiative transfer method \citep[RT, see reviews by][]{Kylafis06,Buat15}.
This method is based on an explicit calculation of the radiation
fields heating the dust \citep{Popescu13}, consequently derived from the attenuated stellar
populations \citep{Tuffs04} in the galaxy under study. In addition, this method uses
constraints provided by available optical information like morphology,
disk-to-bulge ratio, disk inclination (when a disk morphology is present) and
size. 

Since RT methods are notorious for being computationally very time consuming,
until recently they were used only for detailed calculations of a small 
number of galaxies. With the new developments resulting in the creation of 
libraries of RT models, they are now used to derive SFRs in statistical
samples. To this end SFRs were derived by \cite{Grootes13, Grootes14} using 
the RT model of \cite{Popescu11}. Attenuation corrections were produced on an
object-by-object basis, taking into account both the orientation of the galaxy
and the disk opacity. The disk opacity was determined from the opacity - stellar mass
surface density relation of \cite{Grootes13}, making use of
the GAMA single-Sersic morphological fits of 
\cite{Kelvin12}. In principle once the intrinsic stellar SED is derived 
from de-attenuating the observed data, it is
relatively easy to derive a SFR using a calibration between the luminosity of
the stellar photons from the young stars and their sources. We use the Grootes et al. RT-derived
attenuation corrections and apply them to the newer k-corrected \textsc{lambdrar} GALEX-NUV luminosities, and use a luminosity to SFR calibration used in Section \ref{sec:UV} for a Chabrier IMF:

\begin{equation}
\label{eq:RT_lum}
\mathrm{SFR_{\mathrm{RT}} (M_{\odot}\,yr^{-1})=\dfrac{L_{RT-NUV} (W\,Hz^{-1})}{1.38\times10^{21}}}
\end{equation}

The main sources of error on the RT-derived SFRs are outlined in Grootes et al (submitted), and are discussed in detail in Appendix A.

 \begin{figure}
\begin{center}
\includegraphics[scale=0.58]{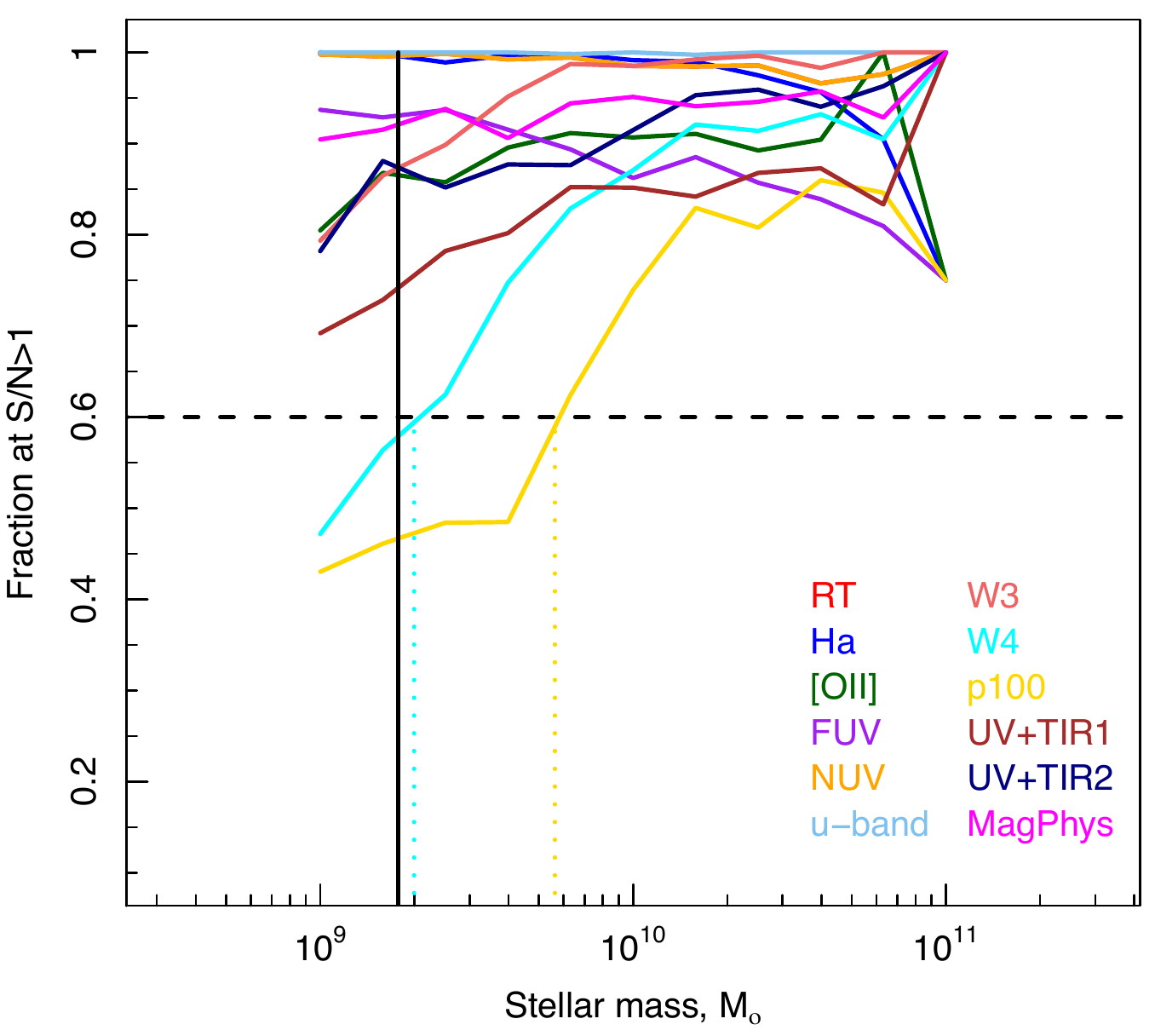}
\caption{The fraction of sources measured at S/N$>$1 as a function of stellar mass, for all SFR methods. We find that the W4 and 100$\mu$m measurements at S/N$>$1 drop significantly at lower stellar masses. In fitting the distributions for these indicators, we apply a sliding lower stellar mass limit defined by the point where the fraction of sources at S/N$>$1 drops below 0.6. These limits at $log_{10}$(M$_{*}$)=9.3 and 9.75 for W4 and 100$\mu$m respectively, shown as the dotted cyan and gold vertical lines. Our sample selection limit of $log_{10}$(M$_{*}$)=9.25 is shown as the vertical solid line. Note that the NUV and FUV measurements have a lower fraction of sources at S/N$>$1 than the RT method, which also uses the NUV measurements. This is due to the fact the NUV and FUV SFRs require a detection in $both$ bands in order to apply obscuration corrections. The medium depth coverage of the GAMA survey footprint is incomplete in the FUV, due to to the failure of the GALEX-FUV detector, and about 30 percent of sources were only covered in the FUV at the shallower depth of the GALEX all-sky survey.}
\label{fig:non_detec}
\end{center}
\end{figure}

\subsection{Comparison of literature SFR methods}
\label{sec:com_ind}

In Figure \ref{fig:SF_vs_SF_old} we display a comparison of a selection of base SFRs for our sample. The relations are fit using the \textsc{[r]} package \textsc{hyperfit}\footnote{http://hyperfit.icrar.org/} \citep{Robotham15}, which also provides a measurement for the scatter orthogonal to the linear best fit - fit parameters are displayed in each of the panels. On each figure, the scatter orthogonal to the linear best fit is a representation of the intrinsic scatter in the distribution, which is not accounted for by measurement error. As such, a small value suggests that the intrinsic scatter in the distribution is much smaller than the scatter in the measurements.

We note that for some indictors a small fraction of un-detected sources have a negative flux measurement (from negative noise measurements at the source position). These sources can not be fit in $log-log$ space. In our fitting process, and all other fitting in this paper, we assign these sources with an error range in $log-log$ space which has its upper boundary defined using the upper error on the linear SFR measurement (which is always positive), and a lower boundary of $log_{10}$(SFR)=-4 (well below the lower range of our SFRs in all indicators). Essentially assuming that these points have a $log$-normal error distribution, we then fit them over this error range.

These un-detected sources do not significantly contribute to the fitted distribution for the majority of SFR methods. However, there are specific stellar mass regimes for particular SFR indicators where non-detected sources dominate the distribution - more specifically, the MIR and FIR measurements at low stellar masses. To highlight this, Figure \ref{fig:non_detec} shows the fraction of sources in our sample which are detected at S/N$>$1 ($i.e.$ have a positive measurement at $1\sigma$ away from zero) as a function of stellar mass. Here we find that the W4 and 100$\mu$m detections drop significantly at low stellar masses, and for the 100$\mu$m, less than $50\%$ of sources are at S/N$>$1 at stellar masses below $log_{10}$(M$_{*}$)$<$9.5. As such, fitting the SFR distribution in these regions does not provide any further information, as the population is dominated by non detections; many with negative flux measurements. To account for this, we apply sliding lower stellar mass limits to the region where we fit the data in all of our analysis for these indicators. This mass limit is defined as the point where the fraction of sources detected at S/N$>$1 drops below 0.6. In practice, this means that we fit the distribution to galaxies down to our original selection limit of $log_{10}$(M$_{*}$)=9.25 in all SFR methods other than W4 and 100$\mu$m - which have a $log_{10}$(M$_{*}$)=9.3 and 9.75 lower stellar mass limit respectively. Note that this only applies to the line fitting, and in our figures we always show the distribution down to the original $log_{10}$(M$_{*}$)=9.25 selection limit.

Considering the resultant fits in Figure \ref{fig:SF_vs_SF_old}, clearly many of these relations have a large scatter ($e.g.$ H$\alpha$ vs NUV), systematic offset ($e.g.$ W3 vs NUV) or incorrect calibration slope ($e.g.$ W3 vs UV+TIR1). These systematic differences in slope and normalisation suggest that the base SFRs are not well calibrated to each other. One of the major aims of this work is to both reduce scatter on these relations (which is not related to measurement error) and remove any systematic differences in slope and normalisation.

\begin{figure*}
\begin{center}
\includegraphics[scale=0.52]{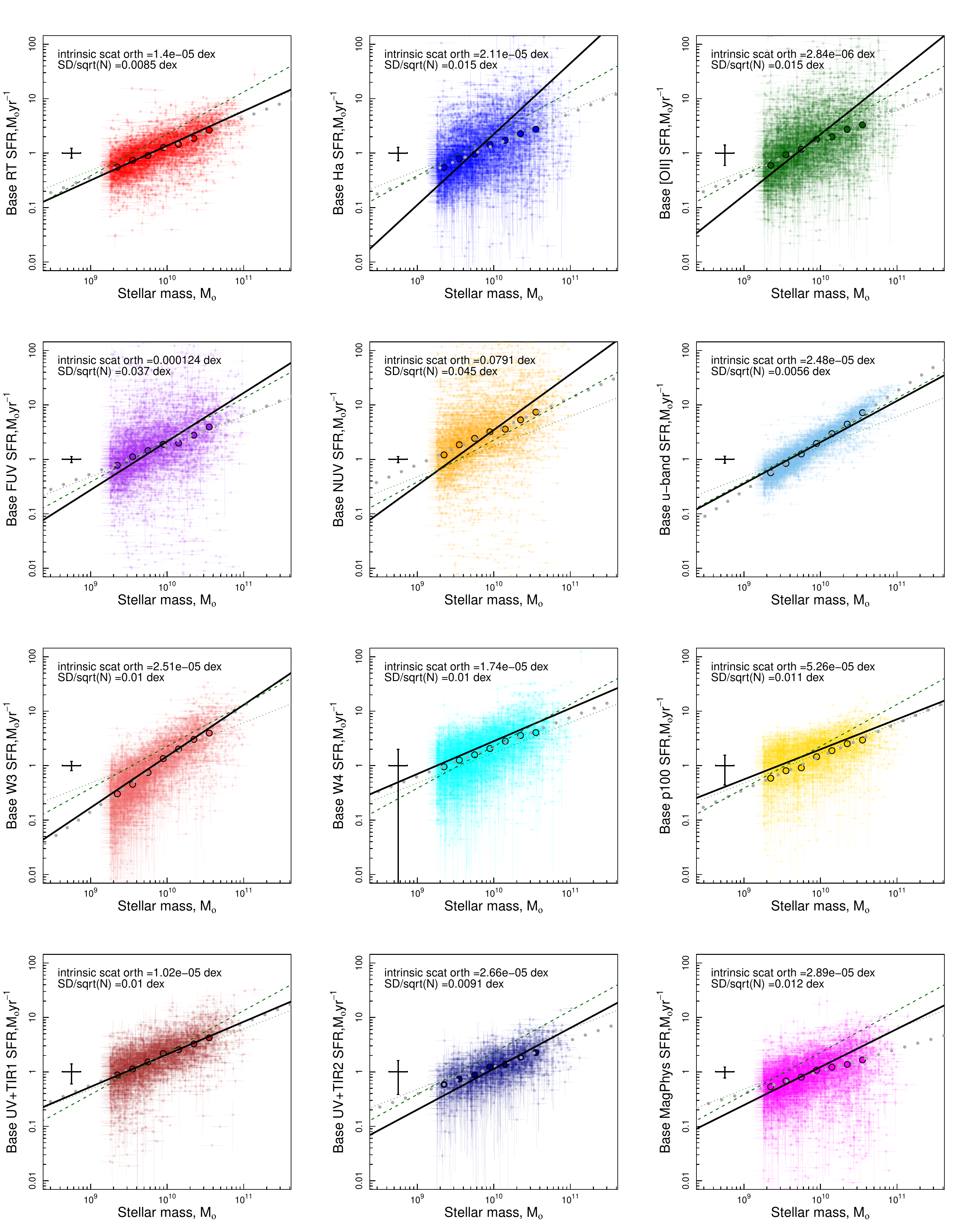}
\caption{The SFR-M$_{*}$ relation for all base SFR indicators available in the GAMA II sample described in Section \ref{sec:indicators}. Note that all figures show an identical sample - that described in Section \ref{sec:data}. The distributions are fit using the \textsc{hyperfit} package. The intrinsic orthogonal scatter (scatter in the distribution on top of measurement error) away from the best fit linear relation are given in each panel and also in Table \ref{tab:calibrations}1. We also display the standard deviation divided by $\sqrt{N}$ at log$_{10}$[M*/M$_{\odot}]=10\pm0.2$ as a measure of the full scatter of the distribution. The green dashed and dotted lines show the SFR-M$_{*}$ relation fits from SDSS at z=0 \citep{Elbaz07} and GAMA I + SDSS at $z<0.1$ \citep{Lara-Lopez13} respectively. A representative median error bar at SFR=1\,M$_{\odot}$\,yr$^{-1}$ is displayed in the left of each panel. The error bar for the base RT panel (top left) is 0.09dex,
which is a lower limit to the total error as estimated in Appendix A. The large circular points in each panel display the running median of the distribution in $log$-$log$ space (to be consistent with our line fitting). The running medians display a largely linear relation between SFR and M$_{*}$ over our sample, for all SFR indicators. To highlight this linearity, we fit the median points with a least squares regression and display the resultant fit as the dark grey dotted line. As such, we deem it appropriate to fit these distributions linearly. }
\label{fig:orig_MS}
\end{center}
\end{figure*}

\section{The Star-Forming Galaxy Main Sequence}

As discussed previously, if a true fundamental relationship derived from the processes by which galaxies form and evolve, the SFR-M$_{*}$ relation should hold true for all SFR measurements, irrespective of method used. While there will be intrinsic scatter in the population (induced by measurement error and variations in SFR timescale), when using a consistent sample, with identical stellar mass estimates, any difference in slope and normalisation between different indicators is likely due to inconsistencies in the calculations of SFR. As such, differences in the slope and normalisation of these distributions can be used as a probe of inconsistencies in the SFR calibrations. As discussed previously, sources with short timescale variations to their star formation history may show scatter in this distribution, but systematic trends are likely to be due to differences in calibration.  

In order to compare and contrast this fundamental relation as viewed by different base SFRs, Figure \ref{fig:orig_MS} displays the SFR-M$_{*}$ relation for all base SFRs discussed in Section \ref{sec:indicators}. Each panel shows an identical sample (described in Section \ref{sec:data}), minus the small number of sources with negative flux measurement, as noted previously. We fit each relation using \textsc{hyperfit} and show the intrinsic orthogonal scatter (in dex) in each of the panels. Some of these scatters are exceptionally small, which highlights that all of the scatter in the data points is consistent with SFR errors, $i.e.$ the intrinsic scatter in the distribution is smaller than the measurement error. A probable exception to this is the scatter in the relation for the SFR derived from NUV using the RT method, which is sufficiently small that a component of the scatter in SFR likely represents true variations in SFR at fixed M$_{*}$,
as discussed in Section 3.9.1. In order to reduce co-variance in the fit errors, we subtract a fiducial mass of M${_f}$=10\,(log[M/M$_{\odot}$]) from the distributions prior to fitting (the SFR fiducial point is already close to zero in log space). Fit parameters to the literature SFR-M$_{*}$ relations, m and C, can be found in Table \ref{tab:calibrations}1 and take the form:

 \begin{equation*}
\mathrm{log_{10}[SFR (M_{\odot}\,yr^{-1})] = m(log_{10}[M*/M_{\odot}]-10)+C}.
\end{equation*}

\noindent We also calculate the standard deviation divided by $\sqrt{N}$ at log$_{10}$[M*/M$_{\odot}]=10\pm0.2$ as a measure of the scatter of each distribution - also displayed in each panel.  

In all of the SFR-M$_{*}$ relation fitting in this work, we opt to fit a linear relation. Many recent works have shown that there is in fact a mass dependant slope to the SFR-M$_{*}$ relation \citep[$e.g.$][]{Whitaker14,Lee15,Schreiber15,Gavazzi15, Tomczak16}. These studies primarily target the slope of the SFR-M$_{*}$ relation at higher redshifts than those probed here ($z>0.5$) and to higher stellar masses (to $log_{10}$(M$_{*}$)$\lesssim$11.5). However, they consistently predict a linear relation (in $log$-$log$ space) at $log_{10}$(M$_{*}$)$\lesssim$10.25, and a flattening to higher stellar masses at their lowest redshifts ($z\sim0.5$) - and this is expected to continue to the more local Universe. Potentially this turnover in the SFR-M$_{*}$ relation is due to different morphological populations contributing to the relation at different stellar masses, with bulge-like systems increasingly flattening the relation at the high stellar mass end. As our sample is selected to target disk-like galaxies only, we potentially may not see this turnover (this will be the subject of an upcoming study). In addition, the bulk of our data lies in the largely linear region of this relation ($log_{10}$(M$_{*}$)$\lesssim$10.25). However, it is nonetheless interesting to explore this further and asses whether it is appropriate to fit our SFR-M$_{*}$ relations linearly. Therefore, in Figure \ref{fig:orig_MS} we also show the running median of each SFR method as a function of stellar mass in $\Delta$log$_{10}$(M/M$_{*}$)=0.2 bins. We then fit the median relation using a least squares regression. In a number of our SFR measurements we do see a suggestion of a flattening of the distribution at higher stellar masses. This is most obvious in the emission-line-derived SFRs (H$\alpha$ and O[II]), but can also be seen in the highest mass bin in the MIR-FIR indicators (W3, W4 and 100$\mu$m). However, in the region where the bulk of our data lies,  $log_{10}$(M$_{*}$)=9.25-10.25 the running medians are well fit by a linear relation in all cases.  As such, we deem it appropriate to fit our SFR-M$_{*}$ relations linearly. Note that the running medians are not always a good representation of the \textsc{hyperfit} fits as they do not consider errors or the spread of the data.

\begin{figure*}
\begin{center}
\includegraphics[scale=0.45]{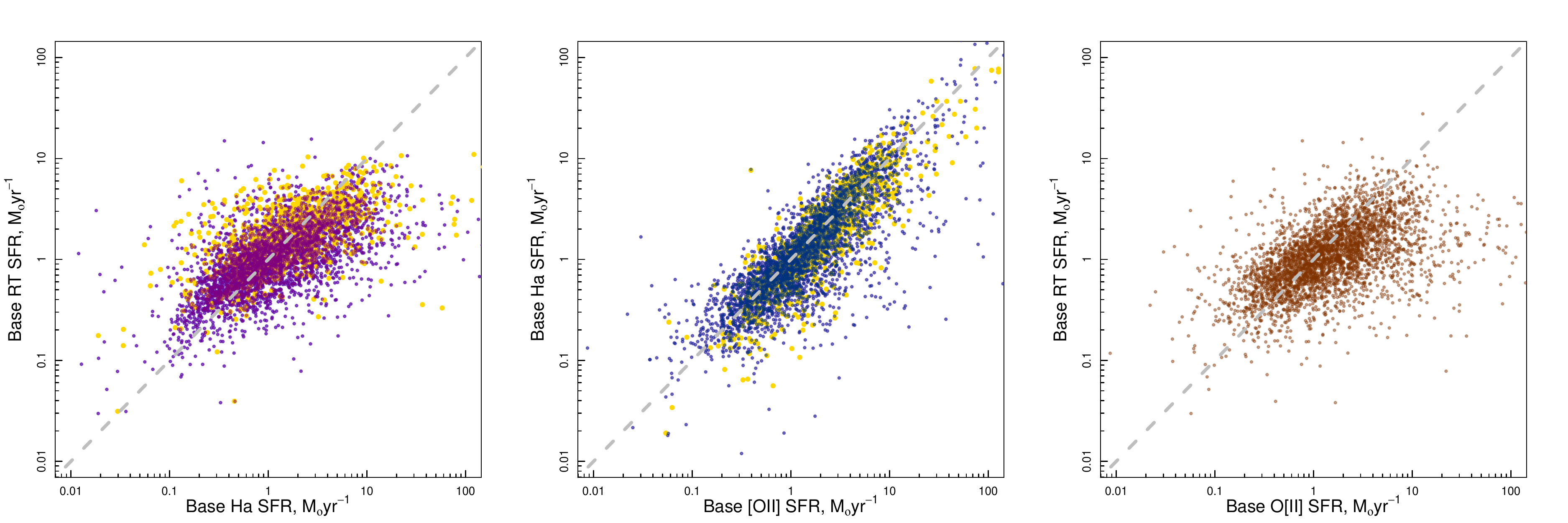}
\caption{Comparison of SFRs derived using the base H$\alpha$ and [OII] calibrations with the RT SFR. We find a tight linear relation between H$\alpha$ and [OII] (middle panel), highlighting that they are probing the same SFR. Both relations are also well calibrated with the RT SFR but deviate slightly in the highest SFR galaxies. For completeness we also show SFRs derived from the observed SDSS spectra (where available), using the same method as the GAMA analysis (gold points) - which are identical to the distribution from the GAMA observations. }
\label{fig:comp}
\end{center}
\end{figure*}

\subsection{Comparison of Methods}
\label{sec:compare}

There are a number of key observations we can derive from Figure \ref{fig:orig_MS} regarding both the physical process involved in producing the emission which is used to derived the SFR and potential errors in calibration:

\vspace{2mm}

\noindent$\bullet$ \textbf{There is relatively large scatter in the H$\alpha$ and [OII] relations}. This is likely to either be due to the timescale over which the emission line SFR indicators probe or the relatively complex corrections applied to emission line-based measurements, such as obscuration, aperture, stellar absorption and/or metallicity corrections. In the former, as discussed previously, H$\alpha$ and [OII] probe SF on very short timescales $<20$\,Myr. As such, they are much more sensitive to recent changes in a galaxy's star formation rate. This short timescale variation in SFR will cause increased scatter in both H$\alpha$ and [OII] derived SFRs. If this is the primary driver of the scatter in the relations, it is not a failing of the calibration, but a true intrinsic property of the galaxy - which can potentially be used to determine a galaxy's recent star formation history. However, if the scatter is induced by the complex corrections, we may require more sophisticated corrections to determine true SFRs from our emission line measurements. This is potentially the case, as aperture corrections are likely to be larger for higher mass galaxies (due to their larger physical size) and the scatter on the H$\alpha$ and [OII] relations increases with stellar mass. By contrast, short stochastic timescale variations in H$\alpha$ and [OII] are likely to be most apparent in lower mass systems - where in fact the scatter is lowest.\\

\noindent$\bullet$ \textbf{The [OII] relation potentially becomes non-linear at $log_{10}$[M*/M$_{\odot}$]$>$10}. There is a suggestion that the [OII] relation turns over at high stellar masses. This is potentially to be due to the saturation of 
metallicity measurements in the high mass regime \citep{Lara-Lopez13}. However, it is promising that the H$\alpha$ and [OII] indictors show a similar distribution in slope and normalisation of the SFR-M$_{*}$ relation, and similar scatter - suggesting that they are both measuring the same physical processes. We explore the relationship between H$\alpha$ and [OII] SFRs further. Figure \ref{fig:comp} displays a direct comparison of H$\alpha$, [OII] and RT SFRs. Clearly the H$\alpha$ and [OII] SFRs are well calibrated and probe similar SFRs - potentially both probing short timescale variations. We also display the SDSS spectra-derived SFRs (using the GAMA analysis of SDSS spectra) which show similar slope, normalisation and scatter, suggesting that the observed scatter is not due to errors in the observed GAMA spectral flux calibration.  Lastly, we also find that both emission line diagnostics are relatively well calibrated with the RT SFR, with a slight normalisation offset.  \\

\noindent$\bullet$ \textbf{The FUV and NUV relations have a large scatter}. This is likely to be due to assumptions in the obscuration correction applied to the FUV/NUV SFRs. For these indicators we only apply a relatively simple and general $\beta_{UV}$ correction, which is likely to be inappropriate for a large number of sources. This is directly apparent when comparing to the base RT SFR, which uses the same NUV photometry but a vastly more sophisticated treatment of obscuration.     \\

\noindent$\bullet$ \textbf{The W3 relation has an offset slope with respect to all other broadband photometric relations}. W3 and W4 are the only broad-band SFRs which are directly calibrated from the GAMA I H$\alpha$ SFRs, as such the slope is likely to be more consistent with the H$\alpha$ relation (as observed in W3). However, this is not apparent for W4. Potentially the W3 calibration has an in built systematic offset which is most dominant at the low mass end, as we will see such offsets will be removed during the analysis in this paper. Another possible explanation is a decline in the strength of PAH features (which is a function of both the strength of the UV radiation field and also of the abundance of PAH molecules) in lower mass galaxies driving the steep slope and large errors at the low mass end. A reduction in PAH abundance can be caused either chemical effects ($e.g.$ a lowering in C/H abundance versus stellar mass) or physical effects ($e.g.$ a more intense and harder UV interstellar radiation field in low mass disk galaxies as compared to high mass disk galaxies - increasing the rate PAH are destroyed in the interstellar medium). This can potentially be explored further using the GAMA dataset.   \\

\noindent$\bullet$ \textbf{The RT, $u$-band, W4, p100, UV+TIR1, UV+TIR2 and} \textsc{magphys} \textbf{relations all show a similar distribution with slightly varying normalisation and slope}. This highlights that even prior to calibration these base SFRs are largely consistent and return similar SFR estimates across a broad range of stellar masses. Small offsets in slope and normalisation will be removed during our subsequent analysis. \\

\noindent$\bullet$ \textbf{The $u$-band relation is linear and has a very small scatter}. This is potentially due to the fact that the $u$-band luminosity is more closely correlated with stellar mass (and not SFR) than the other indicators. However, given the relation between $u$-band SFR and RT/H$\alpha$ SFRs in Figure \ref{fig:SF_vs_SF_old}, it is clear that $u$-band emission is also correlated with SF. As discussed previously, it is likely that the observed $u$-band luminosity has both a SF component and a component from older stellar populations (with produces a correlation with stellar mass). However, we have aimed to remove this from our $u$-band luminosity. \\

\section{Consistent SFRs across all indicators}

\subsection{Deriving new calibrations}
\label{sec:new_cal}

\begin{figure*}
\begin{center}
\includegraphics[scale=0.6]{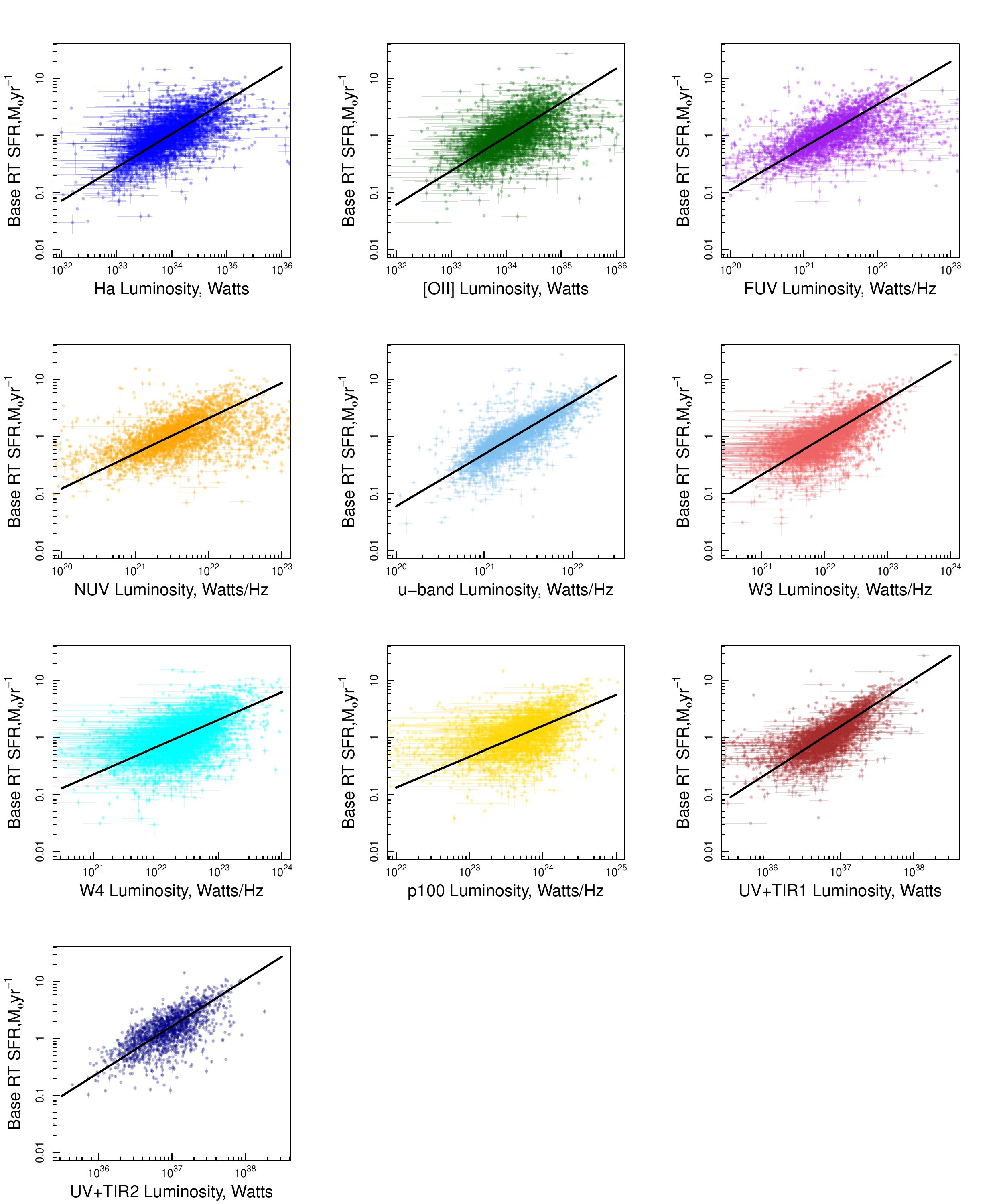}
\caption{The relationship between observed luminosity and base RT star-formation. The distributions are fit using the \textsc{hyperfit} package. We repeat this process using all base SFR calibrations. The best-fit linear relations for all luminosities and base SFR methods, are given in Table \ref{tab:calibrations2}2.}
\label{fig:new_lum}
\end{center}
\end{figure*}

We now aim to recalibrate all of our SFR measurements to form robust and consistent SFR estimates across all methods. These calibrations can then be used to i) compare measurements of SFRs of heterogeneous
samples of galaxies where each galaxy may be measured in a different SFR
indicator and ii) extract the recent SFH of individual galaxies by comparing SFR indicators sensitive to star-formation histories on different timescales. In order to do this we use the following process:
\\
\\
\noindent$\bullet$ Select a base SFR discussed in Section \ref{sec:indicators}.\\
\\
\noindent$\bullet$ Assume this base SFR is representative of the true SFR in the galaxy.\\
\\ 
\noindent$\bullet$ Scale all luminosity measurements to common units of Watts for luminosities and Watts\,Hz$^{-1}$ for luminosity densities. \\
\\
\noindent$\bullet$ For a given base SFR, investigate the relationship between the luminosity measurement used in each method, and that base SFR.\\
\\ 
\noindent$\bullet$ Derive a new best fit linear relationship between luminosity and SFR.\\
\\
\noindent$\bullet$ Repeat this process for all base SFRs.\\

\begin{figure*}
\begin{center}
\includegraphics[scale=0.55]{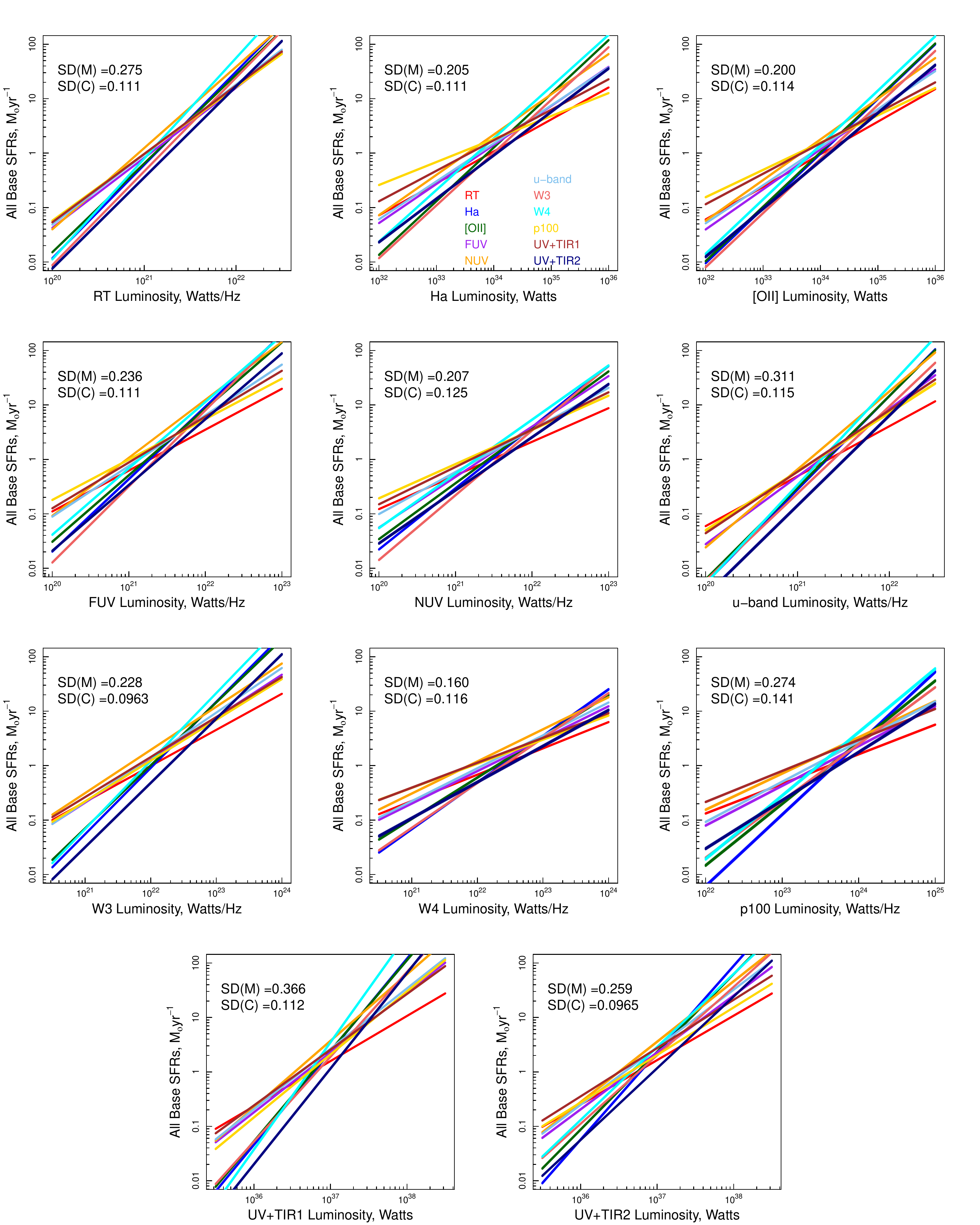}
\caption{Compendium of all observed luminosity to base SFR calibrations given in Table \ref{tab:calibrations2}2. In each panel we take a single observable luminosity, and show the \textsc{hyperfit} derived relation between this luminosity and all literature SFR indicators given in Section \ref{sec:indicators} (displayed by the different colours). For example, the lines fit in Figure \ref{fig:new_lum} for the Grootes et al. NUV literature SFR calibrations are the red lines in each panel of this figure.}
\label{fig:lum_fits}
\end{center}
\end{figure*}

\noindent In this manner we produce a luminosity to SFR calibration for every combination of luminosity measurement, calibrated to each base SFR. The parameters for all new calibrations are given in Table \ref{tab:calibrations2}2. For ease of description, henceforth we only show figures for the RT SFR method in the main body of this paper. Figure \ref{fig:new_lum} displays the observed luminosity for all indicators against the base RT SFR. We once again fit the distributions using the \textsc{hyperfit} package to derive the best fit linear relation - where scatter represents the excess intrinsic scatter not including the known error measurements. Note that we exclude the RT luminosity, as it is our base SFR measurement and we also exclude \textsc{magphys}, as we have no direct luminosity measurement with which to calibrate against. 

Table \ref{tab:calibrations2}2 shows the parameters of the \textsc{hyperfit} fits for all combinations of luminosities and base SFRs. Using this table, it is possible to convert between all SFR indicators using your desired base SFR. For example, if we wish to compare NUV derived SFRs to 100$\mu$m derived SFRs, we can simply re-cast the NUV SFR using the 100$\mu$m calibration or vice-versa. For our final recommended calibrations, please refer to Section \ref{sec:Recom} and Table \ref{tab:pars}. 

In Figure \ref{fig:lum_fits} we show all luminosity-measurement-to-base-SFR relations from Table \ref{tab:calibrations2}2. Each panel displays the relations for a single luminosity measurement, when calibrated against each of the base SFRs. Colours represent each base SFR and are given in the top middle panel ($i.e.$ the top left panel displays the relationship between the RT luminosity and all base SFRs using the calibrators given in Section \ref{sec:indicators} - for reference the fits in Figure \ref{fig:new_lum} are now represented by the red lines in each panel of Figure \ref{fig:lum_fits}). We also display the standard deviation of the slopes (SD[m]) and intercepts (SD[C]) for all fits - as such, a large spread in fit slope across base SFR indicators is represented by a combination of these values. Note that the $x$-axis range in Figure \ref{fig:lum_fits} is not the same for all luminosity measurements (with ranges defined by the spread of the data), such that a small value of SD[m] may not be apparent when visually comparing these figures. The SD[m] and SD[C] parameters display how robust a particular observable luminosity is against measuring SFRs using different methods ($e.g.$ comparing W3 luminosity to SFRs derived using all other base methods, produces reasonably consistent results with $\sim$0.1\,dex scatter).

\begin{figure*}
\begin{center}
\includegraphics[scale=0.58]{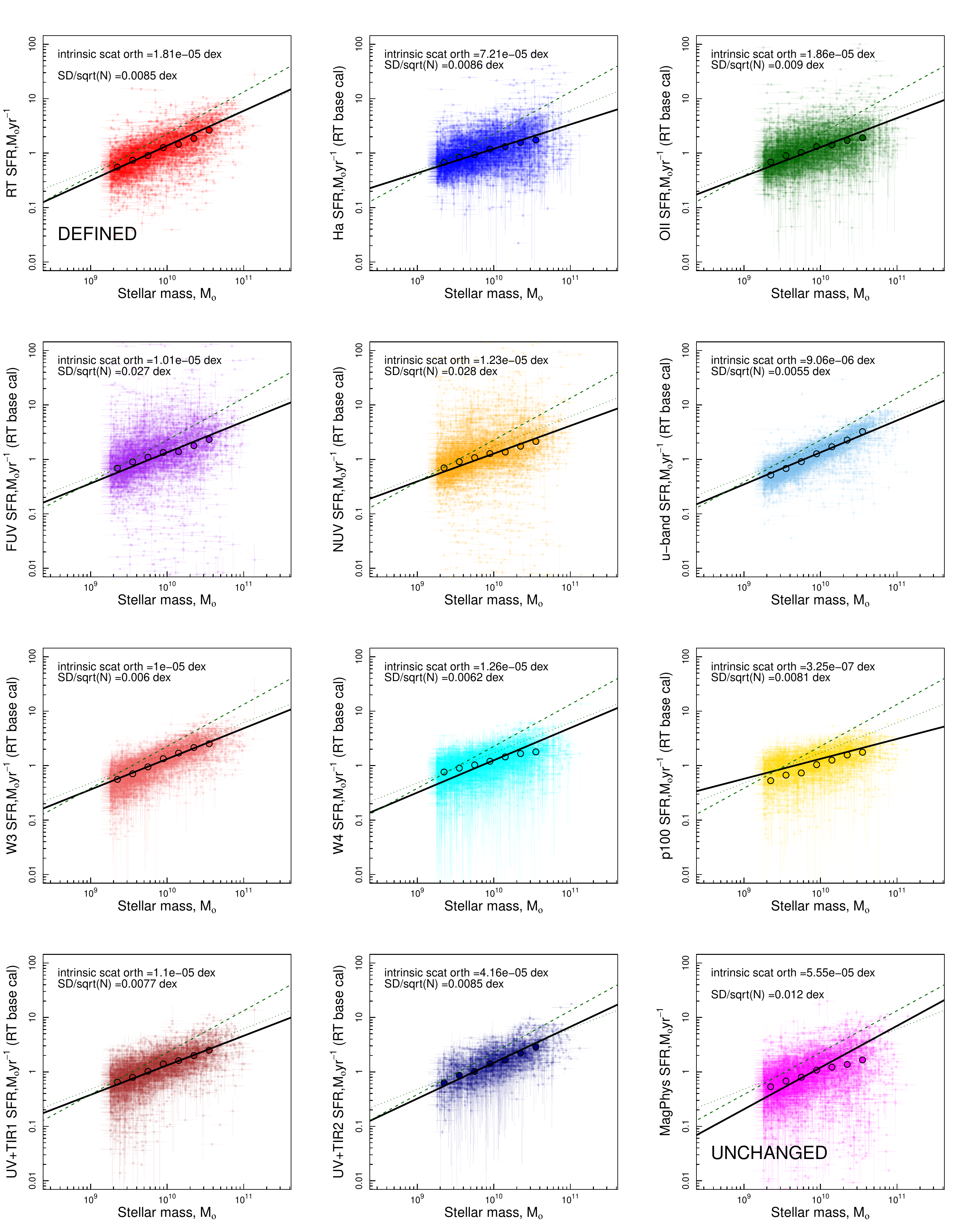}
\caption{The new SFR-M$_{*}$ relation derived by calibrating all SFR indictors using the base RT SFR. This figure takes all of the red lines in Figure \ref{fig:lum_fits} (same as the black lines in Figure \ref{fig:new_lum}), calculates a SFR for each observable luminosity in our sample using these calibrations and reproduces the SFR-M$_{*}$ relation. All orthogonal scatters are improved over the original SFR-M$_{*}$ relations shown in Figure \ref{fig:orig_MS}. The green dashed and dotted show the SFR-M$_{*}$ relation fits from SDSS at z=0 \citep{Elbaz07} and GAMA I + SDSS at $z<0.1$ \citep{Lara-Lopez13} respectively. Once again the large circular points in each panel display the running median of the distribution in $log$-$log$ space}.
\label{fig:new_MS}
\end{center}
\end{figure*}

\subsection{Re-casting the SFR-M$_{*}$  relation}
\label{sec:new_MS}

Using our new luminosity to SFR calibrations defined above, we now derive the SFR-M$_{*}$ relations, recalibrated using each base SFR ($i.e.$ using all combinations of fits given in Table \ref{tab:calibrations2}2). We:
\\
\\
\noindent$\bullet$ Take all of the luminosity to SFR calibrations in Figure \ref{fig:lum_fits}.\\
\\
\noindent$\bullet$ For each luminosity measurement, calculate a new SFR based on these calibrations (this essentially gives us $>100$ SFR estimates for each galaxy) \\
\\
\noindent$\bullet$ Produce a series of SFR-M$_{*}$ relation plots calibrated using each base SFR. \\
\\
For example, for the RT SFR we use all of the red lines in Figure \ref{fig:lum_fits} and for W4 all of the blue lines. An example of the new SFR-M$_{*}$ relation using base RT SFR for recalibration is shown in Figure \ref{fig:new_MS}. We find that using the RT base SFR to recalibrate produces a reduced orthogonal scatter for all SFR-M$_{*}$ relations (this is in fact true for the majority of base calibrators). Full parameters of all new SFR-M$_{*}$ relations are given in Table \ref{tab:calibrations}1.

\begin{figure*}
\begin{center}
\includegraphics[scale=0.43]{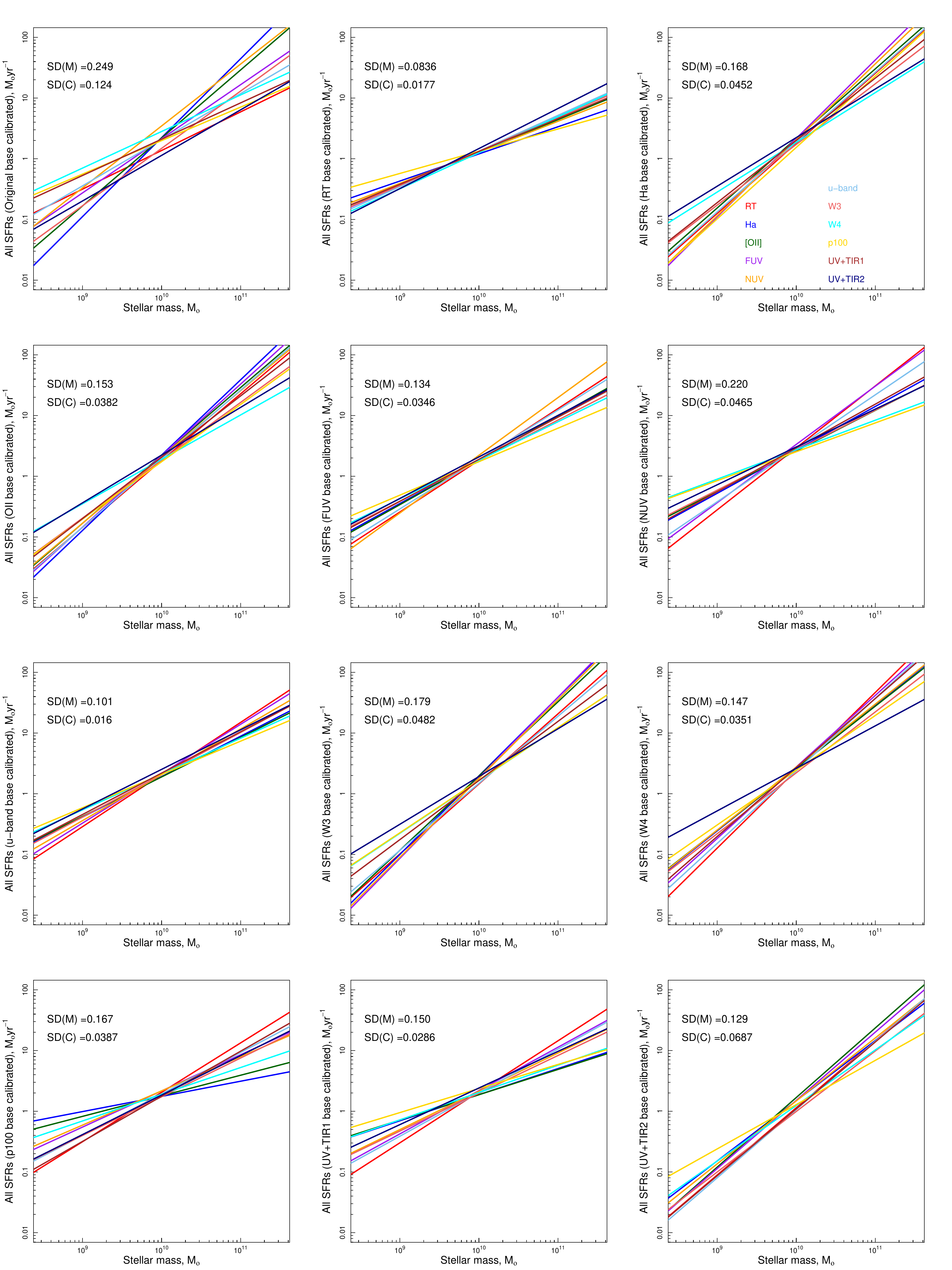}
\caption{Compendium of the SFR-M$_{*}$ relation fits for all combinations of observed luminosity and re-calibrated SFR. Each panel shows the SFR-M$_{*}$ relations derived for all SFR indicators, when calibrated using the literature SFR on the y-axis. As such, the bottom middle panel shows the SFR-M$_{*}$ fits for all indicators (colours) when recalibrated using the literature UV+TIR1 SFR. In the top left we see the original SFR-M$_{*}$ relation fits just using the literature SFRs with no recalibration. Spread on the slope and normalisation are found to be reduced in comparison to the original fits using all recalibrations.  }
\label{fig:MS_fits}
\end{center}
\end{figure*}

In a similar manner to Figure \ref{fig:lum_fits}, Figure \ref{fig:MS_fits} displays all of our new recalibrated SFR-M$_{*}$ fits. Each panel now shows the SFR-M$_{*}$ fits for all SFR methods, re-calibrated using the base SFR tracer on the y-axis. For example, the RT panel (middle top row) shows all of the solid black lines from Figure \ref{fig:new_MS} collapsed to a single figure. We also display the original fits using the base SFR with no re-calibration (top left panel). In order to give a measure of the difference in SFR-M$_{*}$ relations across each indicator, we highlight the standard deviation in both slope and normalisation of all of the fits. 

Comparing all of our new calibrations to the original base SFR-M$_{*}$ fits (top left panel of Figure \ref{fig:MS_fits}), we reduce the spread in slope and normalisation of the fits across all base SFRs. This highlights that we are now measuring SFRs in a much more consistent manner across all indicators and that our recalibrating process is improving consistency when measuring SFRs. It should be noted here that even though SFRs are measured more consistently across different methods for a given base SFR, this in general does not imply that the fidelity of a particular SFR method (corrected according to a different base SFR) has been improved. This will only be the case if the base SFR indicator itself can be shown to
be a superior predictor of the true SFR in a galaxy.  For example, the small spreads in the W3 luminosity corresponding to SFRs of all base methods means that although W3 will give a highly reproducible SFR measurement when used in conjunction with different SFRs, it will still have a systematic error, as discussed in
Section \ref{sec:compare}.

\section{Recommended Luminosity to SFR Calibrations}
\label{sec:Recom}

Here we aim to define a robust set of recommended luminosity to SFR calibrations. To do this we must select one of our base SFRs as a `gold standard', with which to calibrate all methods against. This `gold standard' calibrator must meet the following criteria:
\\
\vspace{2mm}

\noindent 1.\textbf{ The method has a direct, and quantitative link with young massive stars in galaxies.} This criterion is met by indicators which probe either non-ionising continuum UV emitted short-wards of $\sim$3000\AA (RT, FUV, NUV, UV+TIR1/2 and \textsc{magphys}) or ionising photon flux as probed with recombination lines (H$\alpha$ or [OII]). Care must be taken when using broad-band emission at wavelengths $>$ 3000\AA\, ($u$-band) where correlations in the SFR-M$_{*}$ relation can be driven by the fact that flux from older stellar populations may dominate the emission, and this is well correlated with stellar mass. This is discussed further is Section \ref{sec:choice}. 
\\
\vspace{2mm}

\noindent 2.\textbf{ The method must probe integrated star-formation over the whole galaxy.} For the GAMA sample there are no measurements of the spatially integrated H$\alpha$ and [OII] recombination line fluxes as aperture-based measurements are obtained in a $\sim$2\,arcsec aperture which misses large fractions of the line flux, depending on the angular size of the galaxy. While aperture corrections can be made, these are uncertain and may give rise to a scatter in the derived SFR. 
\\
\vspace{2mm}

\noindent 3.\textbf{ The method must not be sensitive to short timescale variations in SFR.} As discussed previously, emission line SFR methods are subjects to short timescale variations in SFR. Such variations will induce scatter in calibrations using these indicators as a `gold standard' measurement. Following this, and point 2., we exclude H$\alpha$ and [OII] as our potential `gold standard' SFR.
\\
\vspace{2mm}

\begin{figure*}
\begin{center}
\includegraphics[scale=0.6]{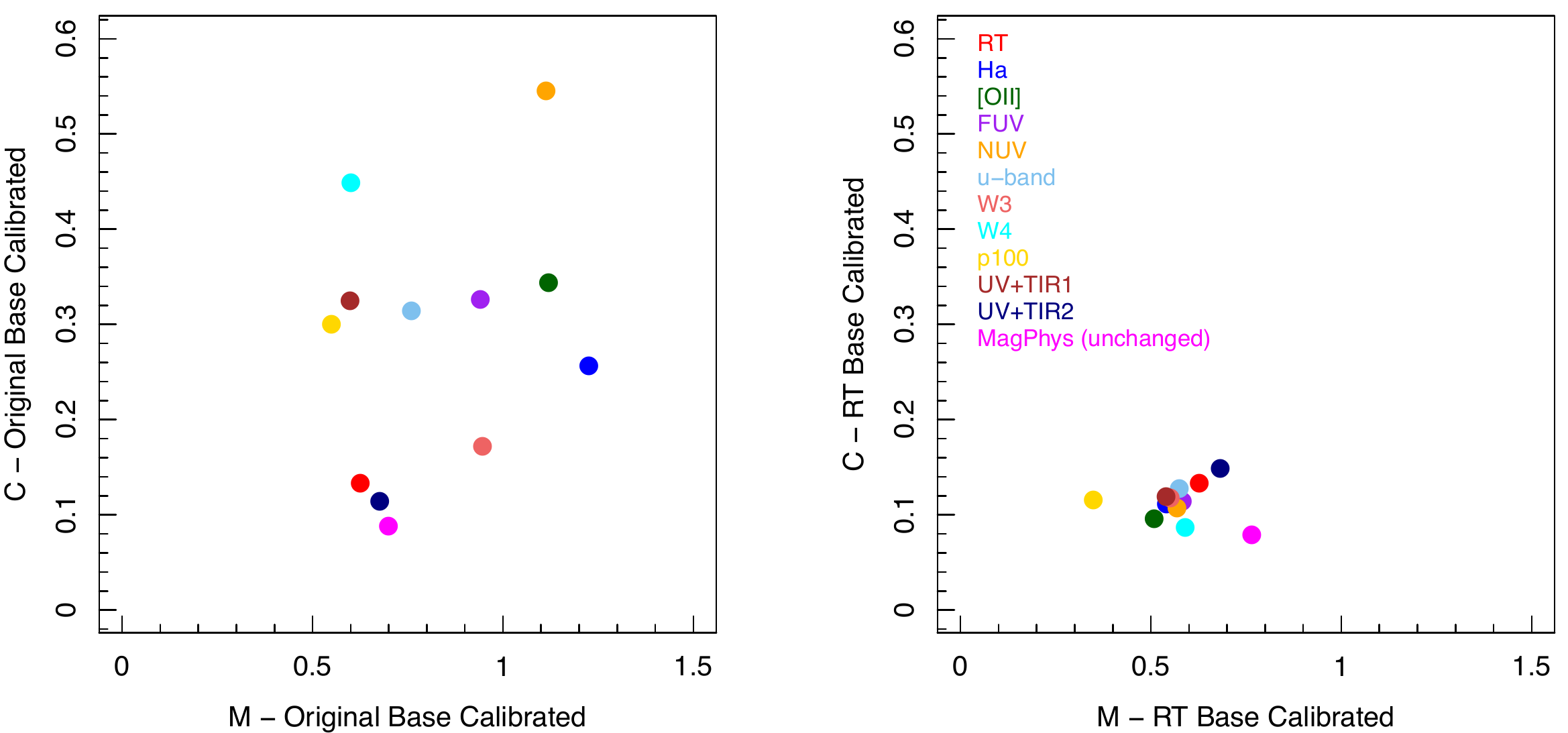}
\caption{The slope, m, against intercept, c of our SFR-M$_{*}$ ration fits using the original SFRs given in Section \ref{sec:indicators} (left) and recalibrated using the RT SFRs as a base indicator (right). These values are representative of the fits displayed in the upper left and upper middle panels of Figure \ref{fig:MS_fits}. We see a large reduction in the spread of both m and c via our recalibration process.}
\label{fig:mvc}
\end{center}
\end{figure*}

\noindent 4.\textbf{ After recalibration using the `gold standard', it must show consistent slope and normalisation in the resultant SFR-M$_{*}$ relations.} Through inspection of the panels in Figure \ref{fig:MS_fits} one sees clearly that using the RT SFR as the base method yields tighter distributions in both slope, m, and intercept, c, than any other SFR indicator used as the base. To highlight the reduction in spread of both m and c when using the RT SFR as our base calibrator, Figure \ref{fig:mvc} displays m vs c for the original SFR-M$_{*}$ ration fits (left) and our new fits recalibrated using the RT SFRs (right). We see a large reduction in the spread of both m and c, highlighting that using the RT method as a base calibrator produces consistent SFR-M$_{*}$ relations.

\vspace{2mm}

Given the above criteria, we select the RT base SFR as our `gold standard'. This measurement is directly linked to the emission from young stars, is integrated over the whole galaxy, is measured over timescales which are unaffected by recent changes to the galaxies star-formation history, has a direct measurement for all galaxies in our sample and after recalibration, produces SFR-M$_{*}$ relations with reduced scatter, and consistent slopes and normalisation. 

It should be noted that the use of the spread in slopes and normalisations of the relations in Figure \ref{fig:MS_fits} as a quantitative measure of the efficacy of the various methods used to derive SFRs (point 4, above), will always preferentially favour methods leading to a relation between log(SFR) and log(M$_{*}$) that has the same linear functional form assumed by \textsc{hyperfit} for the sample of spiral galaxies we use. In this sense our result that the RT method reduces the spread in slopes and normalisations more than the other methods is consistent with the analysis of the SFR-M$_{*}$ relation between specific SFR and
M$_{*}$ by Grootes et al. (submitted), also using the RT technique. That analysis revealed an almost perfect power law relation between 10$^{9-11}$\,M$_{\odot}$ for the same sample of disk-dominated spiral galaxies as used in the present work. As discussed in Section 4 and by Grootes et al. (submitted), one reason for the linearity
may be that this is a fundamental linear relation pertaining only to the disk component of spiral galaxies. The sample selection using the morphological proxies selects against early type spirals, with an increased fraction of bulge light affecting M$_{*}$ but contributing little to the UV output of the galaxies. Thus, the 
possible steepening of the SFR-M$_{*}$ relation at high mass measured in other samples \cite[see for example][]{Whitaker14} may potentially be due to
the increasing predominance of high mass spheroidal components in more massive spiral galaxies.

Another major factor influencing the linearity of the relation between log(SFR) and log(M$_{*}$) is the changing effect of dust attenuation as a function of stellar mass in spiral galaxies. This affects both the escape fraction of UV light as well as the appearance of dust-reradiated optical and UV starlight. These both depend on the geometrical distributions of
stars and dust in the galaxy, as well as the amount of dust, as a function of M$_{*}$. As a result, the transformation between raw measurements in any SFR indicator, and the true SFRs is not predicted to be well described by any simple functional form. The apparent high fidelity of the \cite{Popescu11} radiation transfer model used here in predicting the variation of dust
attenuation in the NUV as a function of M$_{*}$ along the SFR-M$_{*}$ relation, is consistent with the model itself and its associated empirical constraints, having
predictive power for deattenuating spiral galaxies over all the range of  M$_{*}$ considered here. In particular the \cite{Popescu11} model makes use of
the relative spatial distributions of old and young stellar populations and associated dust components
derived from radiation transfer modelling of nearby highly resolved massive edge-on spiral galaxies \citep{Xilouris97, Xilouris98, Xilouris99, Popescu00, Misiriotis01, Popescu04}. These are combined here with the empirical calibration of face-on dust surface density as a function of stellar mass surface density by \cite{Grootes13},
making use of measurements of disk size and orientation derived from single-Sersic fits to the optical images of each GAMA target galaxy in the sample

\begin{figure*}
\begin{center}
\includegraphics[scale=0.75]{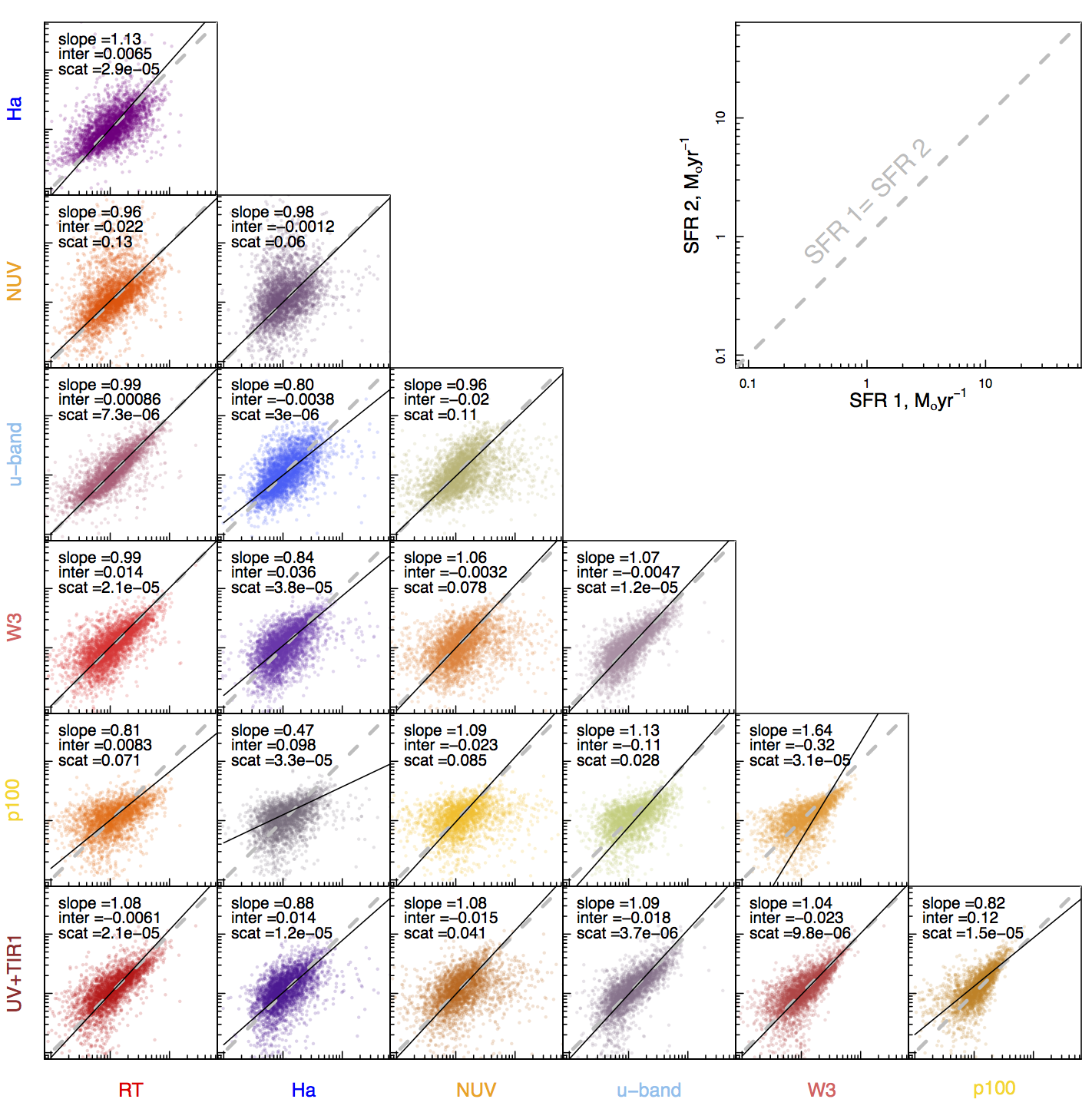}
\caption{Same as Figure \ref{fig:SF_vs_SF_old} but for re-calibrated SFRs using the RT-derived luminosity to SFR calibrations - given in Figure \ref{fig:lum_fits} and Section \ref{sec:new_cal}.  Both axes display SFR from 0.1 to 50 in log-log space. All SFR indicators show a tighter relation which more closely follows the 1:1 line.}
\label{fig:SF_vs_SF_new}
\end{center}
\end{figure*}

In summary, our resultant recommended observed luminosity-to-SFR calibrations are those based on the RT method, and have the form :

\begin{equation*}
\mathrm{log_{10}[SFR (M_{\odot}\,yr^{-1})] = m(log_{10}[L(unit)]-L_{f})+C},
\end{equation*}

\noindent where parameters for each indicator are given in Table \ref{tab:pars}.
 
 \begin{table*}
\caption{Recommended Luminoisty-to-SFR calibrations, based on the radiation transfer model of \citet{Popescu11} in conjunction with the opacity - stellar surface mass density
relation of \citet{Grootes13}. The calibrators take the form: $log_{10}[SFR (M_{\odot}\,yr^{-1})] = m(log_{10}[L(unit)]-L_{f})+C$. Where units are: Watts for broad-band luminosity densities (RT, NUV, FUV, u-band, W3, W4, 100$\mu$m), Watts\,Hz$^{-1}$ for integrated lines luminosities (H$\alpha$, [OII]), and ($L_{IR}+ 2.2 L_{UV}$ Watts) for integrated broadband luminosities (UV+TIR1 and UV+TIR1). The quoted error are errors on the line fits, not the spread of the data. The 'obs' column highlights the luminosity measurement used in this calibration and the final column displays the log$_{10}$(M$_{*}$) range over which our calibrations are derived (as noted in Section \ref{sec:data}). }
\begin{center}
\begin{tabular}{c c c c c c}
SFR Indicator & m & C & L$_{f}$ & Obs & Fit mass range\\
\hline

RT  &  1.00 $\pm$ 0.0097  &  0.11 $\pm$ 0.003  &  21.25 & L$_{RT-NUV}$ & 9.25-12.0 \\
Ha  &  0.59 $\pm$ 0.0097  &  0.031 $\pm$ 0.004  &  34 & Eq. \ref{eq:Ha_lum} & 9.25-12.0 \\ \relax
[OII]  &  0.60 $\pm$ 0.012  &  -0.021 $\pm$ 0.0049  &  34 & Eq. \ref{eq:OII_lum} & 9.25-12.0\\
FUV  &  0.75 $\pm$ 0.03  &  0.17 $\pm$ 0.0055  &  21.5 & Eq. \ref{eq:UV_lum} & 9.25-12.0 \\
NUV  &  0.62 $\pm$ 0.013  &  0.014 $\pm$ 0.0045  &  21.5 & Eq. \ref{eq:UV_lum} & 9.25-12.0 \\
u-band  &  0.92 $\pm$ 0.012  &  -0.079 $\pm$ 0.0044  &  21.25  & Eq. \ref{eq:u_lum} & 9.25-12.0 \\
W3  &  0.66 $\pm$ 0.011  &  0.16 $\pm$ 0.0039  &  22.25 & k-corr L$_{W3}$ & 9.25-12.0 \\
W4  &  0.48 $\pm$ 0.013  &  -0.046 $\pm$ 0.006  &  22.25 & k-corr L$_{W4}$ & 9.30-12.0\\
p100  &  0.54 $\pm$ 0.019  &  -0.063 $\pm$ 0.0095  &  23.5 & L$_{100}$ & 9.75-12.0\\
UV+TIR1  &  0.83 $\pm$ 0.013  &  0.20 $\pm$ 0.0047  &  37 & Eq. \ref{eq:UVTIR1_lum} & 9.25-12.0 \\
UV+TIR2  &  0.82 $\pm$ 0.019  &  0.22 $\pm$ 0.0058  &  37 & Eq. \ref{eq:UVTIR2_lum} & 9.25-12.0\\

\end{tabular}
\end{center}
\label{tab:pars}
\end{table*}

We suggest that the reader use these calibrations to define SFRs for a particular galaxy or galaxy sample, and only refer to the numerous calibrations in Table \ref{tab:calibrations2}2 for a specific comparison between two SFR methods. 

In order to highlight the improvements induced by our new calibrations, Figure \ref{fig:SF_vs_SF_new} displays the same as Figure \ref{fig:SF_vs_SF_old}, but now using the SFRs derived using our recommended calibrations. We find that the scatter is reduced in the majority of relations (potentially removing non-measurement errors), but more significantly, we have dramatically reduced differences in normalisation and slope bringing the majority of the comparisons close to an intercept of 0 and slope of 1 (as expected if SFRs are being measured consistently across indicators).  

We have now derived new luminosity to SFR calibrations which produce the SFR-M$_{*}$ consistently across all SFR indicators ($i.e.$ see the fits for our newly calibrated SFR-M$_{*}$ relations in the top middle panel of Figure \ref{fig:MS_fits}). Thus, using these calibrations one can directly compare SFRs measured using any observable and obtain the same physical property of the observed galaxy.

\section{Evolution of Star-formation in GAMA Galaxies}
\label{sec:dis}

\subsection{Choice of SFR tracer and isolating the star-forming population}
\label{sec:choice}

Using our re-calibrated luminosity to SFR relations, it is possible to derive new, robust SFRs for the full GAMA II$_{Eq}$ sample. As such, we apply our new SFR calibrations to all GAMA II$_{Eq}$ sources - excluding those spectroscopically identified as containing an AGN. In this section we will investigate the evolution of star-formation at $z<0.4$ using the $u$-band derived SFR indicator. We select the $u$-band as: i) the $u$-band calibrations discussed previously in this paper give a relatively tight linear relation (specifically with the robust SFR of the RT method - which is only derived for a sub-sample of GAMA, see Figure \ref{fig:SF_vs_SF_new}), ii) we have robust photometric measurements in the $u$-band for a significant fraction of GAMA galaxies across all redshifts, iii) completeness corrections are much easier to estimate for a single band SFR measurement (in comparison to emission line measurements or full SED fits), and iv) de-blending in photometry from the relatively high resolution SDSS data is likely to have less error than measurements derived from other single bands with poorer resolution (such as GALEX, WISE and Herschel). 

\begin{figure*}
\begin{center}
\includegraphics[scale=0.8]{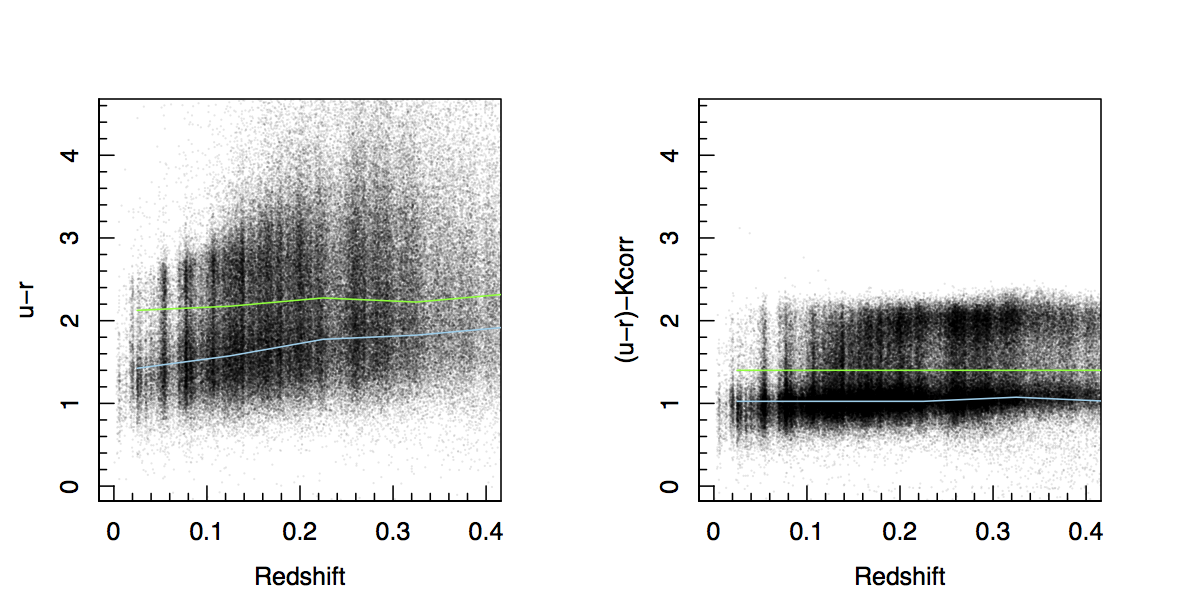}
\caption{The $u-r$ colour distribution as a function of redshift for GAMA galaxies. Left: the measured colours, right: K-corrected and obscuration-corrected colours as used in Section 3. We fit a line to the peak of the blue population (blue lines) and the trough between blue and red populations (green lines). Galaxies which lie below the green line in the right panel are deemed to have their $u$-band emission dominated by young stars and are classed as star forming. In Figure \ref{fig:evolution} galaxies below the green line are shown in colour, while those above the green line are shown in grey.}
\label{fig:col_cut}
\end{center}
\end{figure*}

Furthermore, it is possible to use the radiative transfer models outlined in \cite{Popescu11} and \cite{Grootes13} to predict the most `unbiased' band for estimating the total emission from young stars (and hence SF). We do not discuss this in detail here, as this analysis will be further developed in a subsequent paper. However, such an analysis derives a figure of merit for observing un-biased SF by comparing the fraction of emission arising from young ($<$500\,Myr) stars, to the relative attenuation correction as a function of wavelength. This figure of merit is designed to peak where one most directly measures the emission from young stars, without having to apply significant obscuration corrections. We find that this peaks between the NUV and $u$-bands, suggesting that these bands are optimal for measuring unbiased SFRs.  Given this, and the details noted in the previous paragraph, we select the $u$-band SFRs to measure the evolution of star-formation in GAMA. 

As discussed previously, an additional complication in using broad-band emission to determine global SFRs is that a significant fraction of this emission may arise from the underlying older stellar population, even in star-forming galaxies. However, via our initial colour scaling to the $u$-band luminosity and recalibration process, $u$-band emission which does not arise from star-formation is inherently removed. In the later, this is achieved by directly relating an observed $u$-band luminosity with the RT SFR (which we have assumed to be the true SFR). This allows us to estimate true SFR for a given observed $u$-band luminosity, \textit{irrespective of whether or not that $u$-band luminosity arises from SF}. As we recalibrate in both slope and normalisation, we account for varying fractions of emission arising from older stellar populations at different $u$-band luminosities, with the assumption that this changes linearly.

However, this relation is defined for spiral galaxies (which are predominantly star-formers), and as such, may not be appropriate for passive (largely spheroidal) systems - where the bulk of their $u$-band emission will arise from old stellar populations. While our initial colour scaling should account for this, the corrections for red passive galaxies are large and potentially not robust. As such, to avoid any contribution from non-star-forming components, when estimating the evolution of star-formation in the GAMA II$_{Eq}$ sample we would ideally like to isolate the star-forming population. A potential solution to this is to use emission line derived SFR measurements (which are un-affected by the older stellar populations) as in \cite{Gunawardhana13,Gunawardhana15}. However, this leads to further complications in estimating the appropriate completeness corrections for undetected sources (in addition to aperture and obscuration corrections). Additionally, while it is possible to isolate star-forming and passive galaxies in the SFR-M$_{*}$ plane directly at low redshift, it becomes increasingly difficult at higher redshift where the populations overlap (see below for further details). 

In order to make the distinction between sources where their $u$-band emission is dominated by star-formation and their emission is dominated by older stellar populations we use a simple $u-r$ colour cut as a function of redshift. In Figure \ref{fig:col_cut} we display the measured and k-corrected/obscuration-corrected $u-r$ colours of all GAMA galaxies as a function of redshift. Obscuration-correction and k-correction are once again derived using the \textsc{lambdar}/\textsc{interrest} analysis outlined previously. To separate galaxies into those where their $u$-band emission is dominated by star-formation and those dominated by older stellar populations, we trace the trough point between the blue and red populations (green lines). To do this we bin the $u-r$ colour distribution in $\Delta z=0.05$ bins, determine the peak of the star-forming and passive populations, and find the minimum between these points. We find that the trough point in the k-corrected/obscuration corrected colour distribution is essentially constant with redshift, and opt for a constant $u-r=1.4$ dividing line. We then define all galaxies below the green line in the k-corrected/obscuration corrected colour distribution as `star-forming' in our subsequent analysis. 

For galaxies whose $u$-band luminosity is dominated by recent SF we can use the calibrations derived here for relating luminosity to SFR. However, for those galaxies partially or fully affected by $u$-band emission from older stars (above the green line in Figure \ref{fig:col_cut}) these current calibrations cannot be used because the proportion of $u$-band luminosity attributable to recent SF is not known.

\subsection{Evolution of the SFR-M$_{*}$ relation}   

\begin{figure*}
\begin{center}
\includegraphics[scale=0.46]{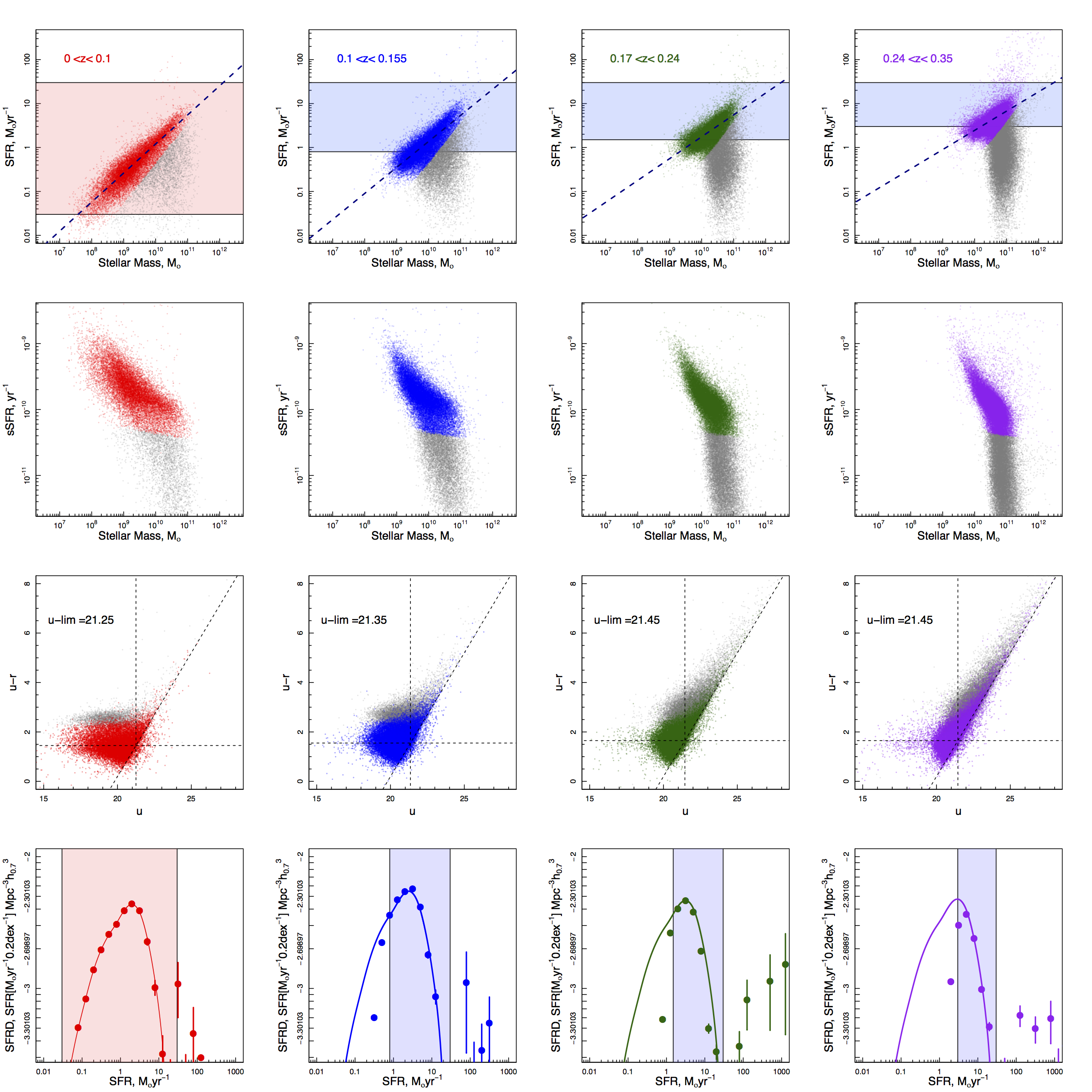}
\caption{Top row: The $u$-band SFR-M$_{*}$ relation for GAMA galaxies in four different redshift bins. Coloured points in all panels display systems selected as `star-forming' using our $u-r$ cuts. Grey points display the passive population, which is not used in subsequent analysis. Coloured points are fit with a linear relation using \textsc{hyperfit} resulting in the navy dashed line - parameters given in Table \ref{tab:slope}. The range of SFRs used in the SFRDF fitting are displayed as the shaded boxes. Second row: The same as the top row but displaying the sSFR-M$_{*}$ relation. Third row: Estimation of incompleteness in each sample. We display the k-corrected (not obscuration corrected as these are not appropriate for observability corrections) $u-r$ vs $u$ distribution of sources at each epoch. The diagonal line displays the GAMA $r<19.8$ limit, the dashed horizontal line displays the modal $u-r$ colour of the 'star-forming' population and the dashed vertical line displays the intercept of these two lines (see text for details). Bottom row: The SFRDF at each epoch. Data points display the 1/V$_{max}$ weighted distributions. We fit the $0<z<0.1$ distribution using a spline and then apply a least squares regression to fit of the same shape to the normalisation and x-position at higher redshifts - using only the data in the shaded regions (once again see text for details).}
\label{fig:evolution}
\end{center}
\end{figure*}

We split the GAMA sample into four redshift bins between $0<z<0.35$. For a direct comparison we use the same redshift binning as in the previous GAMA analysis of \cite{Gunawardhana13,Gunawardhana15}, which uses H$\alpha$ derived SFRs. The top two rows of Figure \ref{fig:evolution} display the SFR-M$_{*}$ (top) and sSFR-M$_{*}$ (middle) relations for galaxies in each redshift bin. We colour those sources which are defined as star-forming using the $u-r$ colour cuts discussed previously. In both rows we see that the distribution of sources visually splits into two populations, and that the simple $u-r$ cut isolates the star-forming population. The sharp cutoff between the coloured and grey distributions is representative of the obscuration and k-corrected $u-r$ colour selection, which is close to a linear cut in sSFR. In the subsequent analysis in this paper we only use this star-forming population, but display the excluded passive population as grey points, where appropriate.   

We use \textsc{hyperfit} to fit the slope and normalisation of star-forming galaxies at each redshift to parametrise the evolution of the SFR-M$_{*}$ relation. Table \ref{tab:slope} displays the best-fit parameters at each redshift.

\begin{table*}
\caption{The evolution of star-formation in GAMA galaxies. We split our sample into four redshift bins and measure the slope and normalisation of the SFR-M$_{*}$ relation. We calculate the SFRD at epoch and estimate its error using the Cosmic Variance (CV) prescription of \citet{Driver10}. }
\begin{center}
\begin{tabular}{c c c c c c}
\hline
Redshift & Volume $\times 10^3$ & CV & Slope & Normalisation ($log_{10}[M*/M_{\odot}]=10$) & SFRD \\
(z) & (Mpc$^{-3}h^{3}_{0.7}$) & (\%) & ($log-log$) & ($log_{10}$[M$_{\odot}$yr$^{-1}$]) & (M$_{\odot}$yr$^{-1}$\,Mpc$^{-3}h^{3}_{0.7}$)\\
\hline

0 - 0.1  &  1339  &  0.13  &  0.66 $\pm$ 0.003  &  0.08 $\pm$ 0.003  &  0.0258 $\pm$ 0.00335 \\ 
0.1 - 0.155  &  3453  &  0.11  &  0.59 $\pm$ 0.004  &  0.14 $\pm$ 0.002  &  0.0409 $\pm$ 0.0045 \\ 
0.17 - 0.24  &  10442  &  0.08  &  0.50 $\pm$ 0.004  &  0.24 $\pm$ 0.002  &  0.0438 $\pm$ 0.00351 \\ 
0.24 - 0.35  &  30872  &  0.05  &  0.44 $\pm$ 0.003  &  0.39 $\pm$ 0.002  &  0.0419 $\pm$ 0.00209 \\ 

\end{tabular}
\end{center}
\label{tab:slope}
\end{table*}

These fits show clear variation in slope and normalisation across our redshift range - with the normalisation increasing and slope appearing to flatten to higher redshifts. To compare this evolution to previous results at higher redshifts, we take the slope and normalisation at log$_{10}$[M/M$_{\odot}$]=10.0 and compare to the parameterisations of the SFR-M$_{*}$ relation given in \cite{Lee15} - using data in the COSMOS region. We select log$_{10}$[M/M$_{\odot}$]=10.0 as our normalisation point, as it falls above the incompleteness limits in all of our redshift samples, and is largely in the linear region of the \cite{Lee15} parameterisations. We then determine the Lee et al normalisation at this point using their equation 2.   

\begin{figure*}
\begin{center}
\includegraphics[scale=0.48]{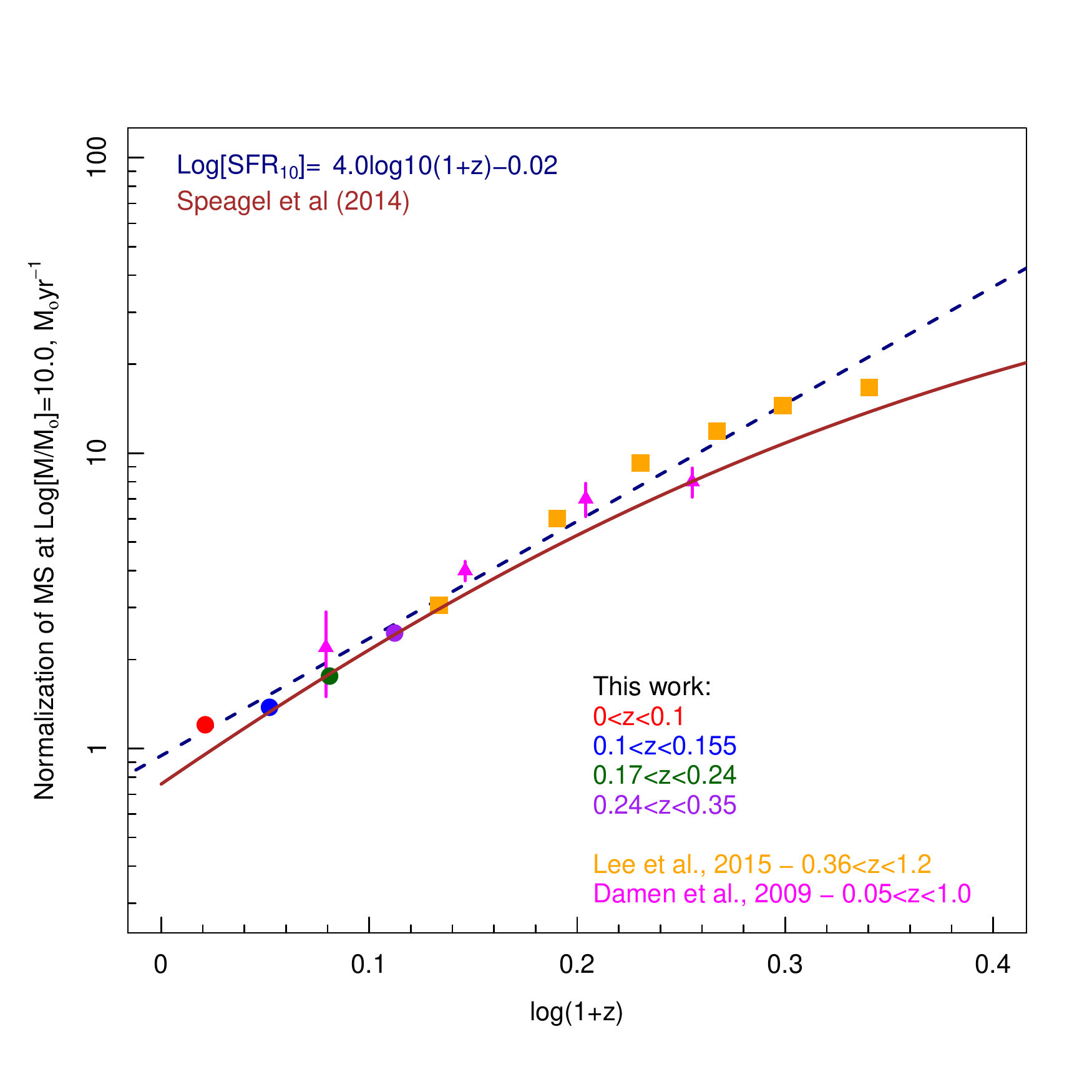}
\includegraphics[scale=0.48]{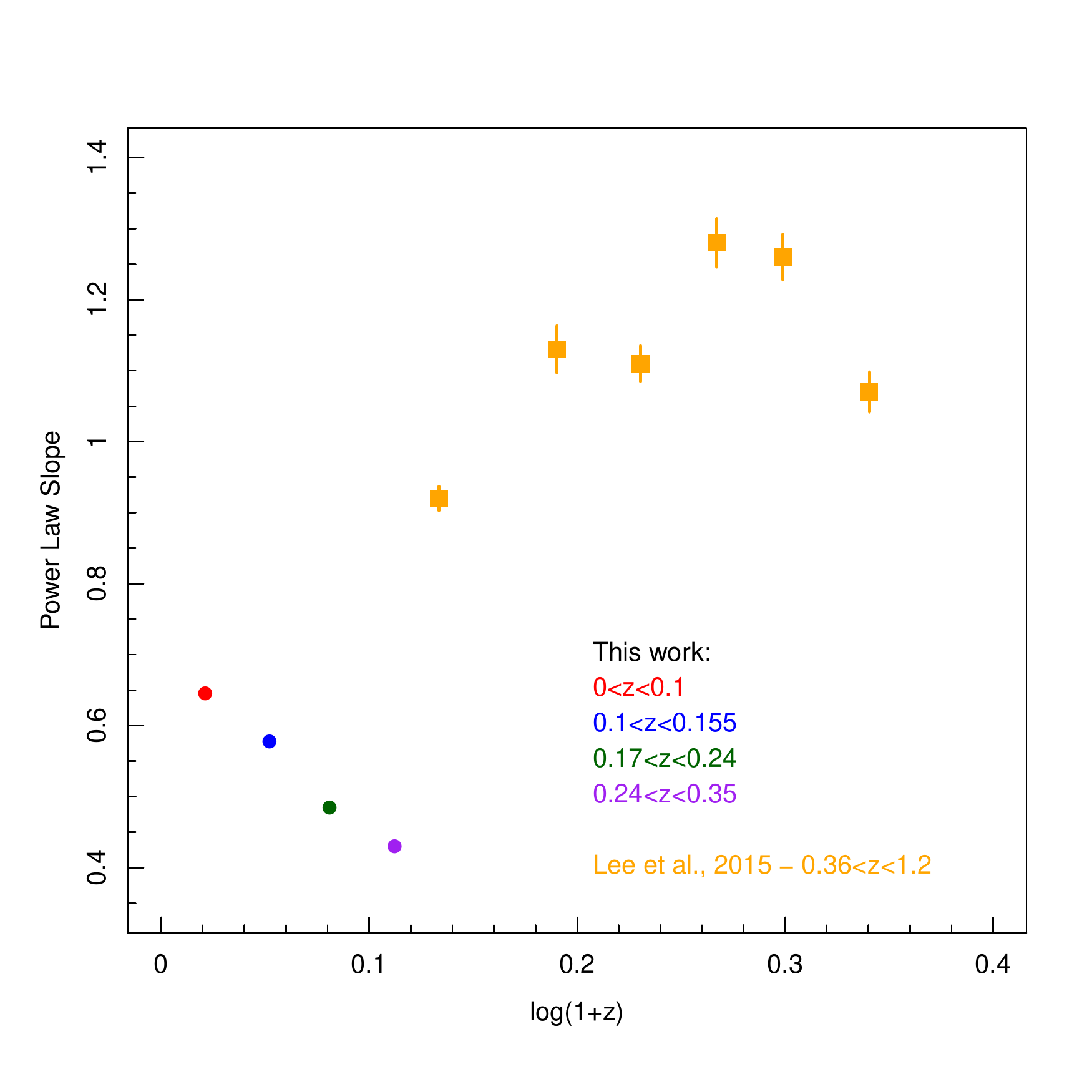}
\caption{Left: Normalisation of the SFR-M$_{*}$ relation evolution at log$_{10}$[M/M$_{\odot}$]=10.0 from our new calibrations in comparison to the \citet{Lee15} results at higher redshifts, observed in the COSMOS region, orange, and points derived from the sSFR evolution in \citet{Damen09}, magenta. The fit using the compilation of data in \citet{Speagle14} is shown as a the brown line. Right: The same but for the power law slope of the SFR-M$_{*}$ relation.}  
\label{fig:normalization}
\end{center}
\end{figure*}

The left panel of Figure \ref{fig:normalization} displays the combined evolution of the normalisation of the SFR-M$_{*}$ relation at log$_{10}$[M/M$_{\odot}$]=10.0 from our work and the \cite{Lee15} study. Errors on the normalisation from our sample are estimated for the \textsc{hyperfit} fitting error and are smaller than the points. We also over-plot points derived from the sSFR evolution in \cite{Damen09} - which are consistent with both our and the \cite{Lee15} results. We find a relatively tight linear relations of the normalisation from $0<z<1.2$, covering the last  8.5\,Gyr of evolution, with the form:

\begin{equation}
\mathrm{log_{10}[SFR_{10} (M_{\odot}\,yr^{-1})] = 4.0log_{10}[1+z]-0.02}
\end{equation}

This highlights that there is relatively smooth declining evolution in the SFR-M$_{*}$ relation since $z\sim1.2$, which is consistent with results from the evolution of the Cosmic Star-Formation History \citep[$e.g.$][]{Hopkins06}, which to first order, has been linearly decreasing since $z\sim1$. 

We also plot the fit for the evolution of the SFR-M$_{*}$ relation from \cite{Speagle14}, who use a detailed compilation of 25 different samples to evaluate the SFR-M$_{*}$ relation out to $z\sim6$. We use their best fit relation for at log$_{10}$[M/M$_{\odot}$]=10.0 and find that our results consistent with the \cite{Speagle14} fits over the redshift range probed.  

We also compare the power law slope of our SFR-M$_{*}$ evolution with that determined by \cite{Lee15} - right panel of Figure \ref{fig:normalization}. While Lee et al propose an increasing slope with look-back time, we in fact find the opposite, with the SFR-M$_{*}$ relation becoming shallower to higher redshift. However, this is potentially due to Eddington Bias \citep{Eddington13}, due to increasing incompleteness in our samples at higher redshifts both in stellar mass and SFR. Similar effects have been discussed in \cite{Dunne09} and \cite{Stringer11}, who find that increasing incompleteness in high-$z$ samples leads to a flattening of the SFR-M$_{*}$ relation. To test this, we take our $z<0.1$ distribution, apply SFR cuts which are similar to the selection limits in the higher redshift bins and re-fit the distribution. We find that the slope of these fits decreases in a similar manner to the evolution of the slope seen at higher redshifts - as such, this is likely to be driven by Eddington Bias. However, deeper, higher redshift spectroscopic samples, such as the upcoming Wide Area VISTA Extragalactic Survey - Deep \citep{Driver15}, will be required to fully probe the changing slope of the SFR-M$_{*}$ relation at these epoch.

\subsection{Evolution of the Cosmic SFR Density}

Taking this further, we can also explore the global evolution of SFR in the local Universe. As discussed previously, the CSFH is a fundamental measurement of the process of galaxy evolution, probing the changing density of star-formation as the Universe evolves \citep[$e.g.$][]{Baldry02}. This distribution contains invaluable information about the underlying processes which shape the evolution of galaxies. The CSFH has been well established over the past two decades from numerous studies, albeit with large scatter from different data sets \citep[see summaries in][]{Hopkins06,Gunawardhana13} - showing an increase in SFR density to $z\sim2.5$ and a slow decrease at later times. \cite{Hopkins06} have produced a robust parameterisation of the CSFH for both Salpeter-like and \cite{Baldry03} IMFs using a compendium of measurements across the last 12.5\,Gyr of Universal history. 

\subsection{Completeness corrections}

Due to both the selection of target sources in the GAMA survey ($r<19.8$) and the $u$-band limits of the SDSS data used to derive our SFR, our star-forming samples are likely to be incomplete, and this incompleteness will vary as a function of redshift. To estimate this incompleteness, we use a similar method to that outlined in \cite{Robotham11a} and \cite{Driver13}. 

Briefly, we first determine the appropriate $u$-band apparent magnitude limit of our samples at each redshift using the $u-r$ vs $u$ colour distributions shown in the third row of Figure \ref{fig:evolution}. Note that the u-r colours displayed in these panels are not obscuration corrected (but are k-corrected) as applying obscuration corrections is not appropriate when determining observability incompleteness. The dashed diagonal line in these figures displays the $r<19.8$ GAMA selection limit. We identify the modal point of the star-forming distribution in $u-r$ colour (dashed horizontal line) and find the intercept of this point with the $r<19.8$ selection limit - giving the appropriate $u$-band limit for each epoch (dashed vertical line). We exclude sources at fainter absolute $u$-band magnitudes than this limit. For all remaining sources we calculate the standard 1/V$_{max}$ correction for both the $r<19.8$ and derived $u$-band limits. Above/below the horizontal dashed $u-r$ line our sample will be incomplete due to the $u$-band/$r$-band limits respectively. As such, we use the appropriate 1/V$_{max}$ for each galaxy depending on its $u-r$ colour, and calculate the resultant star formation rate density function (SFRDF) at each epoch (bottom row of Figure \ref{fig:evolution}). Volumes for each redshift are given in Table \ref{tab:slope}. We find that the SFRDF is bounded in our three lowest redshift bins ($i.e.$ the contribution to the SFR density rises to a peak and then decreases with decreasing SFR), but unbounded at the highest redshifts. This suggests that we cannot reasonably measure the SFR density in our highest redshift bin. Somewhat unsurprisingly, we also find that the dominant contribution to star-formation in the $z<0.15$ universe arises from systems with moderate SFRs ($\sim1-10$\,M$_{\odot}$\,yr$^{-1}$). 

\begin{figure*}
\begin{center}
\includegraphics[scale=0.85]{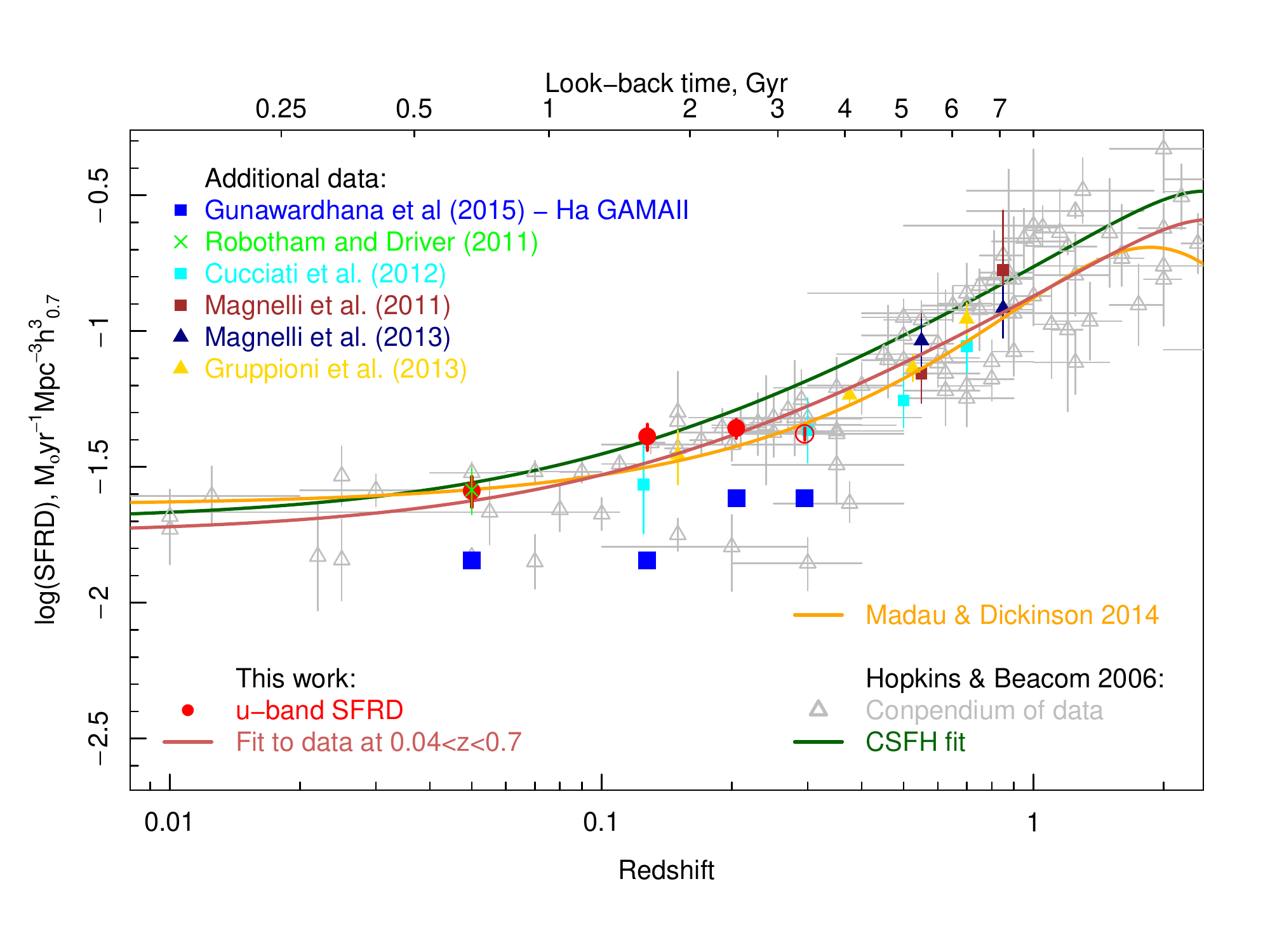}
\caption{The cosmic star formation history. Using our new SFR calibrations, we derive the SFRD in four redshift bins (red circles). Errors are estimated for cosmic variance in the GAMA regions. For comparison we also display the compendium of data from \citet{Hopkins06} - grey, the GAMA-II H$\alpha$ results from \citet{Gunawardhana15} - blue squares,  and results from \citet{Robotham11a} - light green, \citet{Cucciati12} - cyan, \citet{Magnelli11} - Brown, \citet{Magnelli13} - Navy  and \citet{Gruppioni13} - gold. We over plot the fits to the CSFH from both \citet{Hopkins06} - dark green and \citet{Madau14} - orange. We apply a new fit to the CSFH using data in the $0.04<z<0.7$ range and the same functional form as \citet{Hopkins06}, to obtain the salmon coloured line.}
\label{fig:CSFH}
\end{center}
\end{figure*}

While the 1/V$_{max}$ method corrects to weight sources that have been identified, it does not allow correction for sources which are completely undetected in our sample. To account for these sources, we assume that the star formation rate density function has a similar shape at all epochs. We fit the $0<z<0.1$ SFRDF within the bounds of the red shaded region in the bottom left panel of Figure \ref{fig:evolution}, using a spline fit with 8 degrees of freedom (found to be a good representation of the data). We then retain the same shape as the $0<z<0.1$ SFRDF, and apply a least squares fit for the x-offset and normalisation at higher redshifts, using only data points in the region where our SFRDF measurements are complete - the blue shaded regions. For reference, we also over-plot the red and blue shaded regions in the top row of Figure \ref{fig:evolution} to highlight that the fitted region is well within the bounds of the observed galaxy distribution prior to corrections. We note that we aim to fit our highest redshift bin, but highlight that as the distribution is unbounded, any derived measurement is not robust.

\subsection{Cosmic Star Formation History}          

To calculate the Cosmic Star-formation History (CSFH) we integrate each of the SFRDF fits to obtain a total SFRD at each epoch - given in Table \ref{tab:slope}. The dominant source of error in our SFDR measurements is likely to be Cosmic Variance (CV). We estimate the $1\sigma$ error induced by CV in the GAMA regions from \cite{Driver10} as 13\%, 11\%, 8\% and 5\% in our four redshift bins respectively. 

Figure \ref{fig:CSFH} shows a compendium of the CSFH using multiple different surveys and techniques. The grey points display the summary of results noted in \cite{Hopkins06}, which we supplement with more recent results from \cite{Gunawardhana15}, \cite{Robotham11a}, \cite{Cucciati12}, \cite{Magnelli11}, \cite{Magnelli13}  and \cite{Gruppioni13}. Our new GAMA u-band derived SFRD points are shown as red circles - where we highlight our highest redshift bin (where our measurement is not robust) as an open circle. All previous datasets and fits are scaled to a Chabrier IMF.

We over-plot the \cite{Hopkins06} fits and find that our results are largely consistent with the previous parametrisation in the $0<z<0.35$ range - we find that the \cite{Hopkins06} relation slightly over predicts the CSFH at $0.1<z<0.8$ in comparison to our new data and previous results. This is also consistent with the results of \cite{Behroozi13}, who find that the \cite{Hopkins06} relation over predicts the data at these epochs. Lastly, we also display the fit to the CSFH from the summary of \cite{Madau14}, and find it is also largely consistent with our new measurements. For consistency, we also produce the CSFH in an identical manner using the H$\alpha$-calibrated $u$-band SFRs, and find that we obtain SFRD measurements which are $\sim$0.1\,dex higher than our current measurements. Highlighting that our SFRD measurements are not strongly sensitive to choice of `gold standard' calibrator. 

Given the additional data since the \cite{Hopkins06} relation was fit, we now supplement the data used in \cite{Hopkins06} with the more recent results \cite[excluding the][points which are likely to suffer from incompleteness]{Gunawardhana15} and our own new SFRDs (excluding our highest redshift data point), and re-derive the CSFH \citep[for details of the datasets used in the previous fits, please refer to][]{Hopkins06}. We use the same functional form as \cite{Hopkins06}, taken originally from \cite{Cole01} with:

\begin{equation}
SFRD(z)=\frac{(a+bz)h_{0.7}}{(1+(z/c)^d)}, 
\end{equation}

where $h_{0.7}=0.7$. We then perform a least squares regression to fit this relation to all data at $0.04<z<0.7$ (where we have robust measurements of the SFRD) and derive parameters:

\begin{equation}
a=0.0253, \, b=0.167, \, c=3.40, \, d=5.10.
\end{equation}

Note that these fitting parameters are only reliable in the $0.04<z<0.7$ range. We do not quote errors as they are extremely co-variant. Our fitted values do not deviate significantly for the \cite{Hopkins06} relation, but marginally reduce the SFRD at $0.04<z<0.7$ bringing the fitted relation into line with the more recent results.

\section{Conclusions}

We have investigated 12 different SFR methods available to the GAMA survey. By taking a well-defined sample of local spiral galaxies, we compare and contrast each SFR indicator/method against each other and find that they have large scatter, as well as normalisation and slope offsets - suggesting that they are not well calibrated to each other. To produce consistent SFR measurements across all indicators we re-calibrate all SFR metrics to a common relation.

We re-define all literature luminosity to SFR calibrations assuming each base SFR to be correct in turn. We find that the most consistent slopes and normalisations
of the SFR-M* relations found using different SFR indicators are obtained when the indicators are recalibrated using SFRs derived from the radiation transfer method of \cite{Popescu11} in conjunction with the opacity - stellar surface mass density relation of \cite{Grootes13}. We use the SFR-M$_{*}$ relation as a metric of success of our calibrations, and find that our recommended calibrations reduce the scatter on the relation for all SFR indicators and produce consistent SFR-M$_{*}$ relation slopes and normalisation. We also show that our new calibrations reduce scatter and systematic offsets when comparing SFR measurements directly - suggesting that they are robust in directly comparing SFRs across multiple observables. For reference, we also provide luminosity to SFR calibrations for all possible combinations of base SFR and indicator.

These calibrations can now be used to compare SFRs over a broad range of different observations, probing different epochs, environments and galaxy populations, to return a consistent measure of star formation. 

Using our newly derived SFRs we investigate the evolution of star-formation in the local Universe. We apply our new calibrations for $u$-band SFR to the full GAMA II$_{Eq}$ sample and investigate the evolution of the slope and normalisation of the SFR-M$_{*}$ relation. We compare this to results at higher redshift and find that the normalisation evolves linearly over the last 8.5\,Gyr ($0<z<1.2$). We calculate the SFRD at four epochs and correct for incompleteness in our sample. We find that the SFRD derived from our new $u$-band SFR calibrations is largely consistent with the \cite{Hopkins06} and \cite{Madau14} parameterisations of the CSFH. These results highlight that our new SFR calibrations are robust in deriving physical quantities of galaxy populations. 

Lastly, we update the \cite{Hopkins06} relation fits to include post-2006 data, including our new results, to find a slightly lower SFRD at $z<0.7$ - this parameterisation now supersedes the previous relation. Our new relation is also consistent with the more recent results from \cite{Madau14} and \cite{Behroozi13}.

\section*{Acknowledgements}

We would like to thank the anonymous reviewer for their helpful and insightful comments.

GAMA is a joint European-Australasian project based around a spectroscopic campaign using the Anglo-Australian Telescope. The GAMA input catalogue is based on data taken from the Sloan Digital Sky Survey and the UKIRT Infrared Deep Sky Survey. Complementary imaging of the GAMA regions is being obtained by a number of independent survey programs including GALEX MIS, VST KiDS, VISTA VIKING, WISE, Herschel-ATLAS, GMRT and ASKAP providing UV to radio coverage. GAMA is funded by the STFC (UK), the ARC (Australia), the AAO, and the participating institutions. The GAMA website is http://www.gama-survey.org/. 

The Herschel-ATLAS is a project with Herschel, which
is an ESA space observatory with science instruments
provided by European-led Principal Investigator consortia
and with important participation from NASA. The
H-ATLAS website is http://www.h-atlas.org.

MALL acknowledges support from UNAM through the PAPIIT project IA101315. L. Dunne acknowledges support from European Research Council Advanced Investigator grant Cosmic Dust.

\appendix

\section{Error Estimation for Radiative Transfer SFRs}

Considering all aspects of the radiation transfer model of \cite{Popescu11}
we identify three main components to the uncertainty in the SFRs derived
in this application of the model. These are errors related to measurement
errors in disk angular sizes, errors related to
uncertainties in the fixed vertical heights
of the NUV-emitting stars and dust in the geometrical model, and
errors related to the escape fraction of NUV light from star-forming
regions in the disk.
These errors dominate the uncertainties arising from random measurement
errors in the GALEX NUV photometry, for the mass and redshift range
of the galaxies in the sample.

We first consider the effect of
measurement errors in the value for the single-Sersic effective
radius, R$_{eff}$, of the disk. Measurements of R$_{eff}$
are used to convert dust masses into face-on
B-band central optical depth, $\tau_{B}$, following the method of \cite{Grootes13}.
While random errors in R$_{eff}$ arising from noise on
the optical images quoted by Kelvin et al. (2012) are small, we expect
larger systematic errors due to i) the effect of dust on
the perceived optical morphology, ii) the possible presence of a
bulge on single-component Sersic fits of disks, and iii) the effect of
fitting a 2-dimensional analytic function which cannot describe the
projection of the actual 3-dimensional emissivity distribution.
All these systematic effects have been extensively simulated and
documented by \cite{Pastrav13a, Pastrav13b},  and corrected for in our analysis,
as described in \cite{Grootes13}. Nevertheless, residual
errors will remain - for example due to likely variations
in intrinsic bulge-to-disk ratios between galaxies which
cannot be modelled by single-sersic fits. An upper limit to
the effect of these residual errors on $\tau_{B}$
is given by the observed scatter in $\tau_{B}$
about the opacity-stellar surface density
relation of \cite{Grootes13} at high stellar mass densities, which is found to be 
around 0.1dex (1$\sigma$). For a typical galaxy of
stellar mass 2 $\times$ 10$^{10}$M$_{\odot}$ and $\tau_{B}$=3.0,
viewed at the median inclination of
60deg, this translates into a 1$\sigma$ uncertainty in SFR (determined from the NUV) of 0.07dex.

The second main component to the error relates to uncertainties
in the assumed form and parameter values
for the geometrical distribution of dust relative to that
of the UV stellar emissivity in the two disk model of the radiation
transfer method. One such uncertainty
relates to the fact that we are applying an axisymmetric model
which also ignores the structure due to spiral arms. However,
as demonstrated by \cite{Popescu11}, and the comparison
between the 3- and 2-dimensional models by \cite{Natale15},
this appears to have only a relatively small effect on the predicted
escape
fraction of the integrated UV light from spiral galaxies.
More important for the present application are 
uncertainties in the fixed relative scale heights of the
UV-emitting star-forming disk and the dust disks -
which in the model of \cite{Popescu11} are based on resolved
observations of nearby edge-on spiral galaxies by \cite{Xilouris99}
and on the
thickness of the star-forming molecular layer in the Milky Way.
Uncertainties in the scale heights will affect both the attenuation of UV
light in disks seen face on, as well as the inclination dependence
of the attenuation. \cite{Grootes13} has verified that
that the geometry calibrated in this way is
consistent with the observed attenuation-inclination
relation of spiral galaxies in the UV, including the
dependence on stellar mass of this relation. This demonstrates
that systematic errors may be limited. Nevertheless,
random variations in the relative scale heights of the
stellar and dust disks are expected, and these will
contribute to a random error in attenuation, and hence SFR.
If the observed range of
scaleheight of optically emitting stars and associated dust layer
derived through radiation transfer analysis of nearby
highly resolved edge-on spiral galaxies by \cite{Xilouris99}
and by \cite{deGeyter14}
is also indicative of the range of scaleheights of UV-emitting
stars and associated dust, variations of up to a factor of
2 from the mean relative scaleheights of UV-emitting stars to
the associated dust layer may be possible. For a typical galaxy of
stellar mass 2$\times$10$^{10}$M$_{\odot}$ and $\tau_{B}$=3,
such a shift in the vertical geometry
would change the measured NUV flux by $\pm$0.04dex.

The final main component of the uncertainty in SFRs derived using the
RT method relates to uncertainties in the escape fraction of observed
emission in the NUV-band from massive stars arising from the birth clouds.
In the present calculation, the escape fractions as a function
of rest frame UV wavelength are taken from Table E.4 of
\cite{Popescu11}, corresponding to a clumpiness factor
$F$=0.35 in the model. However, it is possible that the clumpiness
factor varies between galaxies, which would induce a scatter in the
returned values for SFR about their true values. Indeed, radiative
transfer
modelling of direct and dust-radiated
light of spiral galaxies by \cite{Misiriotis01} does show a variation
of
$F$ by of order $\pm$30$\%$ between galaxies. This would translate into a
variation in derived SFR of $\pm$0.04dex.

Summing these three components of uncertainties in SFR in quadrature,
yields an estimate for the total uncertainty in the RT method
to be around 0.09dex. This may be a lower limit to the total uncertainty
if other effects are present that we have not taken into account,
but cannot readily quantify, such as deviations in the radial
distribution of dust with galactocentric radius in galaxies
from the exponential functional form used in the \cite{Popescu11}
model. A conservative upper limit to the error
is given by the residual scatter at fixed M$_{*}$
in the sSFR vs. M$_{*}$ for spiral galaxies
of typically 0.27dex (1$\sigma$) found by Grootes et al. (2016, submitted)
after having applied the same RT technique on the same sample of
non-grouped
non-interacting galaxies used in this work. However, the large
difference between the residual scatter of 0.27dex and the
estimated error of 0.09dex suggests that much of the residual scatter
is in fact intrinsic, caused by real variations in the SFR at fixed
M$_{*}$, rather than being due to unaccounted uncertainties in the correction
for dust attenuation.

Finally, it is instructive to compare this lower limit of
0.09dex in uncertainty in SFR
with the reduction in scatter at fixed M$_{*}$ of the
sSFR vs. M$_{*}$ relation for this galaxy sample, as
effected by this implementation of the RT technique.
This reduction is from 0.36dex to 0.27dex
(Grootes et al. 2016 submitted). From this, one may infer
that the component of scatter in SFR at fixed M$_{*}$ in the
uncorrected sSFR vs. M$_{*}$ relation due to dust attenuation
in the NUV is at least 0.24dex for a spiral of stellar
mass 2 $\times$ 10$^{10}$M$_{\odot}$.

\begin{figure*}
\begin{center}
\includegraphics[scale=1]{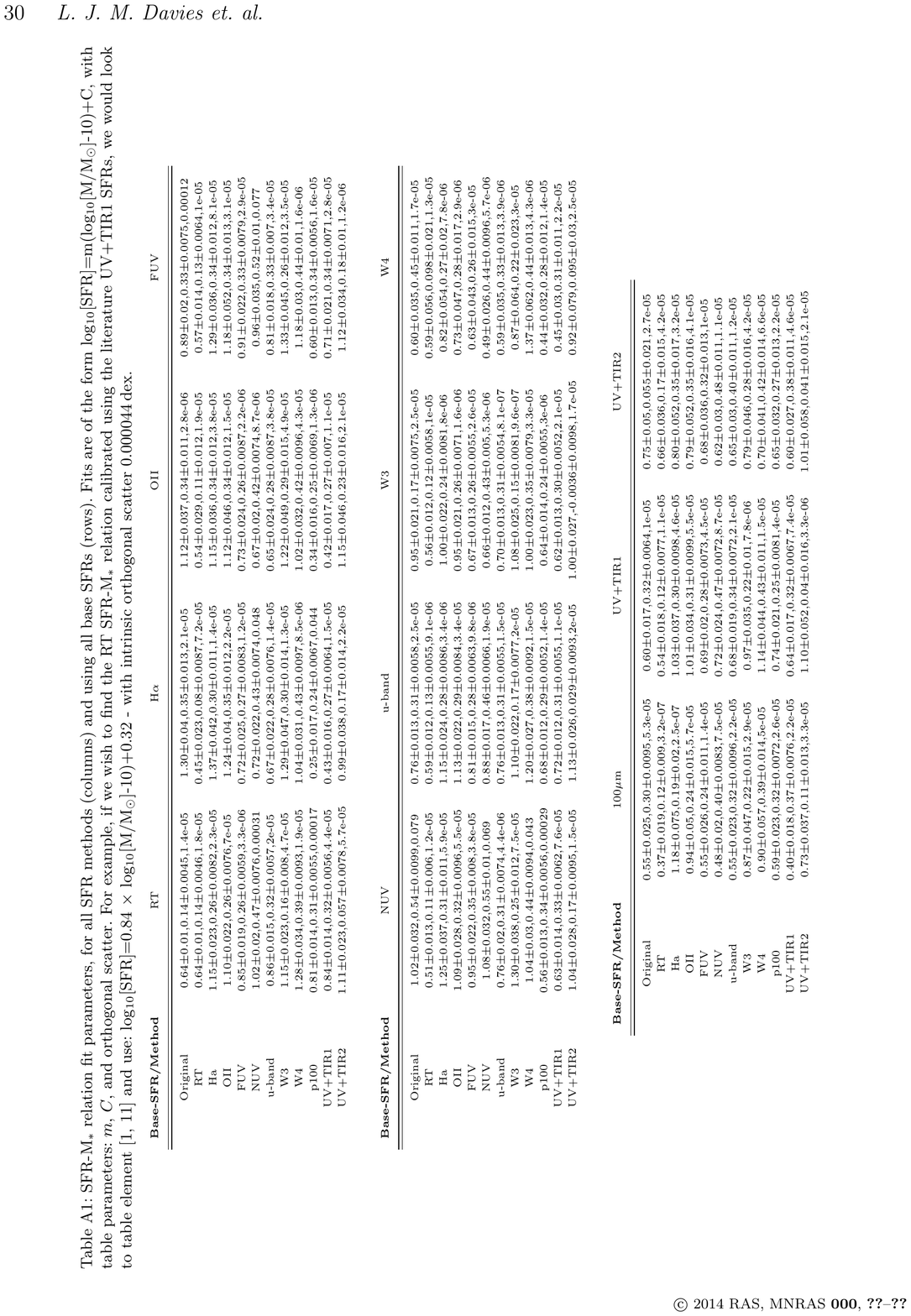}
\label{tab:calibrations}
\end{center}
\end{figure*}

\begin{figure*}
\begin{center}
\includegraphics[scale=1]{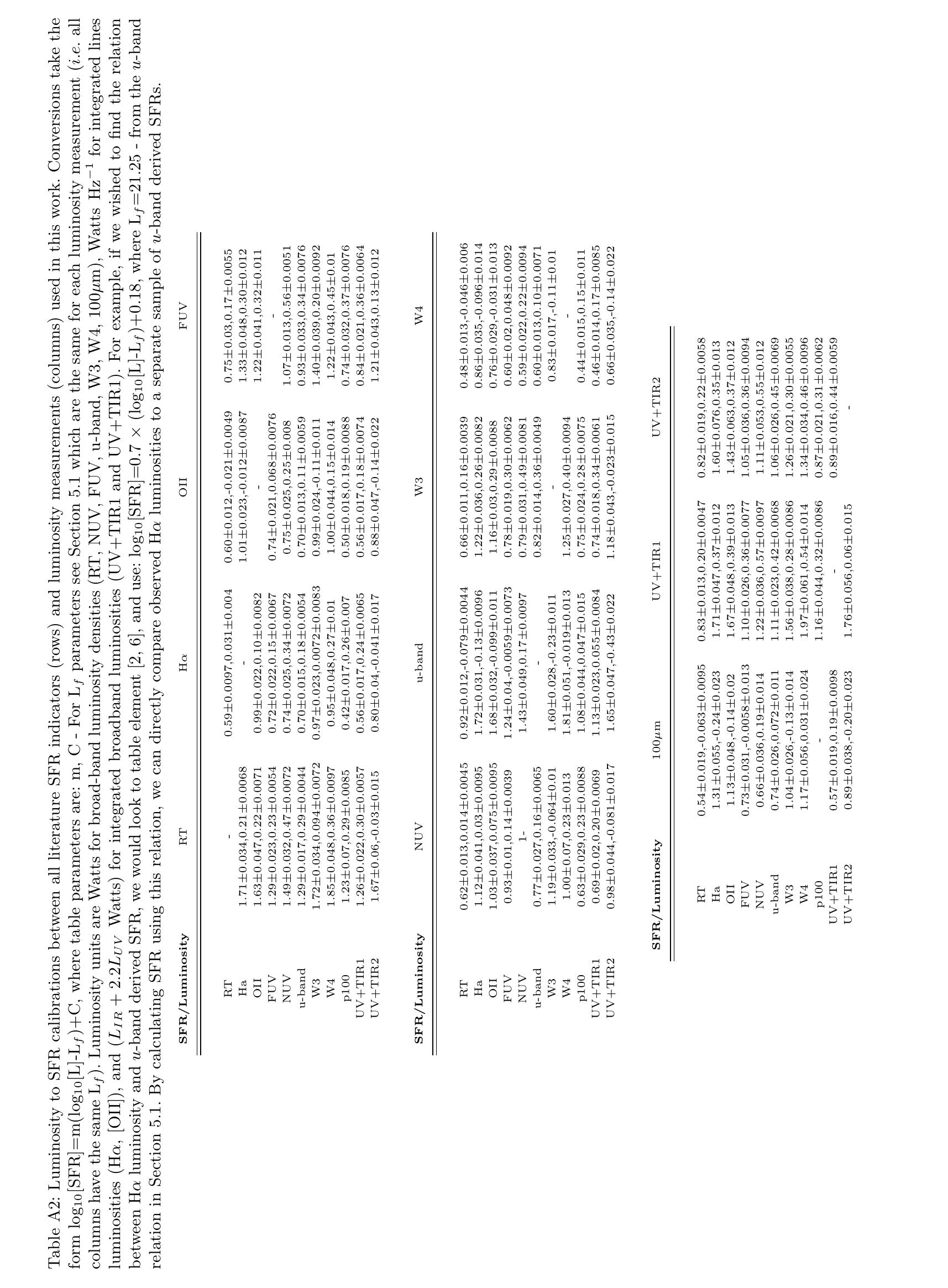}
\label{tab:calibrations2}
\end{center}
\end{figure*}

\end{document}